\documentclass[]{emulateapj}
\usepackage{graphicx, amsmath, amsthm, amssymb}

\def\ba{\begin{eqnarray}}
\def\ea{\end{eqnarray}}
\shorttitle{Stability and fates of hierarchical two-planet
 systems}
\shortauthors{Petrovich}

\begin{document}

\title{ The Stability and fates of hierarchical two-planet systems}
\author{Cristobal Petrovich\altaffilmark{1}}
\altaffiltext{1}{Department of Astrophysical Sciences, Princeton University, Ivy Lane, Princeton, NJ 08544, USA; cpetrovi@princeton.edu}

\begin{abstract}
We study the dynamical stability and fates of 
hierarchical (in semi-major axis) 
two-planet systems with arbitrary eccentricities
and mutual inclinations.
We run a large number of long-term 
numerical integrations and use the Support Vector Machine 
algorithm to search for an empirical boundary that best separates
stable systems from systems experiencing either ejections or
collisions with the star.
We propose the following new criterion for dynamical stability:
$a_{\rm out}(1-e_{\rm out})/\left[a_{\rm in}(1+e_{\rm in})\right]>2.4
\left[\max(\mu_{\rm in},\mu_{\rm out})\right]^{1/3}(a_{\rm out}/a_{\rm in})^{1/2}+1.15$,
which should be applicable to
planet-star mass ratios $\mu_{\rm in},\mu_{\rm out}=10^{-4}-10^{-2}$,
integration times up to $10^8$ orbits of the inner planet, and mutual inclinations
$\lesssim40^\circ$.
Systems that do not satisfy this condition by a margin
of $\gtrsim0.5$ are expected to be unstable,
mostly leading to planet ejections if $\mu_{\rm in}>\mu_{\rm out}$, 
while slightly favoring collisions with the star for $\mu_{\rm in}<\mu_{\rm out}$.
We use our numerical integrations to test other stability criteria
that have been proposed in the literature 
and show that our stability criterion performs significantly better
for the range of system parameters that we have explored.
\end{abstract}                                     
  
\section{Introduction}
\label{sec:intro}

More than $\sim50$ exoplanet systems 
discovered by radial velocity (RV) surveys are known
to harbor at least two planets, and many of them are in
eccentric 
and well-separated 
orbits.
The search for and characterization of these planets
in either RV or transit surveys
is generally a time-consuming task, and 
having an-easy-to-use and accurate dynamical 
stability criterion is important to constrain either the existence of extra
planets in the systems or the orbital configurations
of already confirmed planets.

Another motivation for searching for a stability criterion
comes from theoretical studies in which
 planets can have a variety of fates
depending on the dynamical stability of a 
planetary system.
For example, instability can lead to the formation 
of free-floating planets  through planet ejections 
(e.g., \citealt{sumi11,VR12}) and 
planets reaching very nearly parabolic orbits
can collide with or be tidally disrupted by the host star, 
becoming a possible source of stellar 
metal pollution (e.g., \citealt{sandquist02,zuck03,veras13}).
Similarly,  long-term stable and well-spaced planetary systems can 
evolve secularly (with no orbital energy exchange) to
form close-in planets by high-eccentricity
migration
(e.g., \citealt{naoz11,WL11,teyss13,petro15}). 
A simple criterion to decide the fate of a planetary
system based on its observed orbital configuration
can help to constrain the most likely evolutionary path 
of different exoplanet systems without using expensive
long-term $N$-body experiments.

There is no analytic stability criteria for arbitrary 
eccentricities and/or inclinations (see \citealt{geor08}
for a review), while the currently available
(semi-) empirical criteria 
(e.g., \citealt{H72,EK95,MA01})
have generally not been tested in the planetary regime
(in which one body contains almost all the mass of the system) 
or for the long timescales (up to $\sim10^6-10^8$ orbits) 
during which two-planet systems can still become
unstable \citep{VM13,PTR14}.

In this study, we search for empirical criteria to
decide whether a hierarchical 
two-planet system is likely to remain
stable for long timescales or lead to either ejections or
collisions with the host star.
We extend previous numerical work
(see \S \ref{sec:previous_stab})
by considering a wider
range of planetary systems, with planets in
eccentric and/or mutually inclined orbits, 
and much longer evolution timescales. 
We also use, for the first time, the
Support Vector Algorithm in the context 
of dynamical stability analysis, and fully detail
our implementation.


\section{Previous work on the stability of two-planet systems}
\label{sec:previous}

In this section, we briefly summarize the previous work
on the stability of two-planet systems. We will use some
of the stability criteria that have been proposed
in the literature as benchmarks 
to compare to our results in  \S\ref{sec:performance}.

\subsection{Stability of 
close two-planet systems with low eccentricities}
\label{sec:stab_low}
 If the orbits of two planets are guaranteed to never cross, 
 precluding collisions between 
planets or strong gravitational interactions,
 then they are said to be {\it Hill stable}. 
It has been shown that the conservation of 
angular momentum and energy can constrain 
Hill stable trajectories \citep{MB82,MN83}.

For low eccentricities ($e\lesssim0.1$),
the Hill stability criterion 
can be written as (e.g., \citealt{gladman93})
\ba
\frac{a_{\rm out}}{a_{\rm in}}>2.4 
\left(\mu_{\rm in}+\mu_{\rm out}\right)^{1/3}+1,
\label{eq:hill_low}
\ea 
where $a_{\rm out}$ ($a_{\rm in}$) and 
$\mu_{\rm in}$ ($\mu_{\rm out}$) are the semi-major axis and 
planet-to-star mass ratio of the inner (outer) planets,
respectively.
For reference, Equation (\ref{eq:hill_low}) implies that two Jupiter-like 
planets orbiting a Sun-like star are Hill stable for
$a_{\rm out}\gtrsim1.30 a_{\rm in}$. 
Note that the Hill stability criterion does not
 discriminate mean-motion resonances.

The Hill criterion gives no information about the 
long-term behavior of the system,
and  repeated interactions between planets in Hill stable
orbits  can still lead to either ejections and/or collisions
with the star. 
The orbits that are protected against either ejections
or collisions with the star are referred to as 
{\it Lagrange stable}.
Also, the systems that fail the Hill criterion can
 still avoid having close approaches and be long-term stable
 (see discussion \S\ref{sec:performance}).

While there is no analytic criterion for Lagrange stability,
numerical studies show that the Lagrange
stability boundary lies close to the Hill stability boundary
\citep{BG06,BG07,deck12}. 
Based on the  first-order mean-motion 
resonance overlap criterion 
\citep{wisdom80,duncan89}, \citet{deck13} studied
the conditions that can yield chaotic behavior 
in a two-planet system. 
These authors give the following criterion for
the onset of chaos (which implies instability
in their experiments) for two-planet
systems in circular orbits:
\ba
\frac{a_{\rm out}}{a_{\rm in}}<1.46 
\left(\mu_{\rm in}+\mu_{\rm out}\right)^{2/7}+1.
\label{eq:lag_low}
\ea 
For reference, from this criterion 
two Jupiter-like planets are Lagrange unstable
for $a_{\rm out}\lesssim1.25 a_{\rm in}$.
A numerical refinement of the chaotic zone boundary, which sets
the stability condition above, is provided
by \citet{MM15}.
Also, \citet{VM13} numerically studied the relation between the Hill 
and Lagrange stability boundaries for different 
eccentricities.


Based on this previous work, hierarchical (or well-spaced, 
say $a_{\rm out}\gtrsim2 a_{\rm in}$) and coplanar
two-planet systems with low eccentricities are all expected 
to be long-term stable (e.g., \citealt{marzari14}).
Thus, the question of long-term stability in hierarchical two-planet
systems should be focused on
planets in eccentric orbits.


\subsection{Stability of
hierarchical two-planet systems with arbitrary 
eccentricities}

\label{sec:previous_stab}

There is no analytic criterion for 
the stability of hierarchical two-planet
systems in eccentric orbits\footnote{Extensions 
to the first-order resonance overlap criterion to
eccentric orbits require taking into account higher-order 
mean-motion resonances (e.g., \citealt{deck13}). 
A calculation for non-zero (but still low) eccentricities considering 
only first-order resonances in the test-particle approximation has been
carried out by \citet{MW12}.
} 
and most previous works rely on numerical 
experiments and/or heuristic approaches.
We summarize some of the 
dynamical stability criteria proposed for hierarchical
two-planet systems. 
For consistency, we express
each stability boundary in the form
\ba
r_{\rm ap}\equiv\frac{a_{\rm out}(1-e_{\rm out})}{a_{\rm in}(1+e_{\rm in})}>Y
\label{eq:rap}
\ea
where 
$e_{\rm in}$ ($e_{\rm out}$) is the eccentricity of 
the inner (outer) planet
and $Y$ is a function of 
the initial orbital elements and masses.
This choice is motivated by our results in \S\ref{sec:single_alpha}
where we find that the single parameter that
best describes the stability boundary is $r_{\rm ap}$.

(i) \citet{EK95} studied the stability of hierarchical triple systems 
with a wide range of masses
and define a system to be $n$-stable if it preserves
the initial ordering of the semi-major axes  
of the orbits and there are no escape orbits for 
$10^n$ orbits of the outer planet.
The authors find an empirical condition for
2-stability using a set of N-body integrations, which 
in the planetary regime ($\mu_{\rm in},\mu_{\rm out}\ll1$)  
becomes:
\ba
r_{\rm ap}>
Y_{\rm crit}^{\rm EK95}&\equiv&1+3.7\mu_{\rm out}^{1/3}+\frac{2.2}{1+\mu_{\rm out}^{-1/3}}
+\nonumber \\
&&1.4\mu_{\rm in}^{1/3}\frac{\mu_{\rm out}^{-1/3}-1}{1+\mu_{\rm out}^{-1/3}}.
\label{eq:EK95}
\ea
The authors tested this criterion for $\mu_{\rm in},\mu_{\rm out}\geq0.01$ 
and the following sets of orbital elements: 
prograde coplanar orbits with either $e_{\rm in}\in[0,0.9]$ and $e_{\rm out}=0$
or $e_{\rm out}\in[0,0.9]$ and $e_{\rm in}=0$, and
circular orbits with mutual inclinations $i_m\in[0,180^\circ]$.

(ii) \citet{MA01} 
made an analogy between the
stability against escape in the three-body problem
and the stability against chaotic energy exchange
in the binary-tides problem, and derived 
a semi-analytic stability criterion.
We modify their criterion for coplanar and prograde
orbits to express in the form of Equation (\ref{eq:rap})
as
\ba
r_{\rm ap}>
Y_{\rm crit}^{\rm MA01}\equiv 2.8\frac{(1-e_{\rm out})}{(1+e_{\rm in})}
\left[(1+\mu_{\rm out})\frac{1+e_{\rm out}}{(1-e_{\rm out})^{1/2}}\right]^{2/5}.
\label{eq:MA01}
\ea
This criterion does not include a dependence on
$e_{\rm in}$ and $\mu_{\rm in}$, but the authors claim that it is
valid for all eccentricities and masses of the inner
body. Also, this criterion was proposed in the context of
stellar clusters where, unlike our study, 
the mass ratios are not too different 
from unity.

(iii) The Hill stability criterion by \citet{MB82} can be written 
in the planetary regime as 
\citep{gladman93}:
 \ba
r_{\rm ap}>Y_{\rm crit}^{\rm Hill}\equiv
\delta^2\frac{(1-e_{\rm out})}{(1+e_{\rm in})},
\label{eq:Y_hill}
\ea
 where $\delta$ satisfies the implicit equation
 \ba
&&\frac{\left(\mu_{\rm in} + \mu_{\rm out}/\delta^2\right)}{\left(\mu_{\rm in} + \mu_{\rm out}\right)^3}
\left[\mu_{\rm in}\left(1 - e_{\rm in}^2 \right)^{1/2}+
\mu_{\rm out}\left(1 - e_{\rm out} \right)^{1/2}\delta\right]^2\nonumber\\
&&-1-3^{4/3}\frac{\mu_{\rm in}\mu_{\rm out}}{\left(\mu_{\rm in}+\mu_{\rm out}\right)^{4/3}}
=0.
\ea
The Hill stability condition has been extended 
by  \citet{VA04} and \citet{donnison06,donnison11} to arbitrary mutual
inclinations $i_{\rm m}$.
Even though the Hill stability might not determine the
long-term stability of a two-planet system  (see \S\ref{sec:stab_low}),
we will use it as a benchmark \citep{BG06,BG07}.
 \citet{KB10} numerically studied the relation between 
 Hill and Lagrange stability 
 and provided fitting expressions to determine the relation
 between these boundaries. Their results should be 
 applicable to planetary systems consisting of one 
 terrestrial-mass planet and one much more massive planet
 with initial eccentricities less than 0.6. 
 In this work, we focus on a complementary regime in which 
 both the inner and the outer planets have masses much 
(at least $\sim30$ times) larger than the Earth and therefore
do not attempt to compare our results with those 
 by  \citet{KB10}.
 
(iv) \citet{GMC13} proposed a semi-empirical
stability criterion for eccentric two-planet systems
based on Wisdom's criterion of first-order mean-motion
resonance overlap \citep{wisdom80}. 
The authors argue that the initial value of the 
relative longitudes of pericenter
$\Delta\varpi=\omega_{\rm in}+\Omega_{\rm in}-(\omega_{\rm out}
+\Omega_{\rm out})$
can have a significant effect on the stability boundary,
where $\omega_{\rm in}$ ($\omega_{\rm out}$) and
$\Omega_{\rm in}$ ($\Omega_{\rm out}$) are the argument of pericenter
and the longitude of the ascending node of the inner (outer) orbits, 
respectively.
For the most conservative case 
with $\Delta\varpi=180^\circ$ the stability boundary is given
by
\ba
r_{\rm ap}>
Y_{\rm crit}^{\rm GMC13}\equiv 1+1.57\left[\mu_{\rm in}^{2/7}
+\mu_{\rm out}^{2/7}
\left(\frac{a_{\rm out}}{a_{\rm in}}\right)\right].
\label{eq:GMC13}
\ea
The authors also provide expressions for the 
case in which the ellipses are initially aligned
($\Delta\varpi=0$), but an expression for arbitrary
values of $\Delta\varpi$ is not provided.

\begin{table*}[ht]
\begin{center}
\caption{Summary of simulated systems and outcomes}
\setlength{\tabcolsep}{2.5pt}
\renewcommand{\arraystretch}{1.7}
    \begin{tabular}{c|ccccc|c|cccc}
\hline
\hline
Name&$a_{\rm out}/a_{\rm in}$ & $e_{\rm in}$, $e_{\rm out}$ &inc. &$\mu_{\rm in},\mu_{\rm out}$ 
&$t_{\rm max}$ &$N_{\rm sys}$ &2 pl. with & 2 pl. with & Ejection  & Stellar Coll.\\
&&&[deg]&[$M_J/M_\odot$]&[$P_{{\rm in},i}$] &&$\left|\frac{\Delta a}{a_i}\right|<0.1$& 
$\left|\frac{\Delta a}{a_i}\right|>0.1$&
$\left(\frac{r}{a_{{\rm in},i}}>10^2\right)$&$\left(\frac{R_\star}{a_{{\rm in},i}}=10^{-4}\right)$\\
\hline
{\it 2pl-fiducial}&U(x; 3,10) &U(x; 0,0.9) &Ray(1)& U(log x; -1,1) & $10^8$&3567& 2319& 8 & 911 &329   \\
\hline
{\it 2pl-fid-4}&U(x; 3,10) &U(x; 0,0.9) &Ray(1)& U(log x; -1,1) & $10^4$&3567& 2917& 390 & 212 &48   \\
{\it 2pl-fid-5}&U(x; 3,10) &U(x; 0,0.9) &Ray(1)& U(log x; -1,1) & $10^5$&3567&  2672& 258 & 482 &155   \\
{\it 2pl-fid-6}&U(x; 3,10) &U(x; 0,0.9) &Ray(1)& U(log x; -1,1) & $10^6$&3567& 2480& 80 & 747 &260   \\
{\it 2pl-fid-7}&U(x; 3,10) &U(x; 0,0.9) &Ray(1)& U(log x; -1,1) & $10^7$&3567& 2369 & 16 & 874 &308   \\
\hline
{\it 2pl-inc-0}&U(x; 3,10) &U(x; 0,0.9) &1& U(log x; -1,1) & $10^7$&2000 & 1389& 31 &422 &158   \\
{\it 2pl-inc-20}&U(x; 3,10) &U(x; 0,0.9) &20& U(log x; -1,1) & $10^7$&2000 & 1422& 12 &402 &158   \\
{\it 2pl-inc-rand}&U(x; 3,10) &U(x; 0,0.9) &U(x; 0,80) & U(log x; -1,1) & $10^7$&5000 & 3319& 37 &1144 &500   \\
\hline
   \multicolumn{11}{l}{ \footnotesize {\bf  Note.} $P_{\rm{in},i}$ is the initial period of the inner planet. $ U(x; x_{\rm min}, x_{\rm max})$ is the uniform distribution with $x_{\rm min}<x<x_{\rm max}$ and $\mbox{Ray}(x)$}\\
      \multicolumn{11}{l}{is the Rayleigh distribution with parameter $x$.}
\end{tabular}
\end{center}
\label{table:all_sim}
\end{table*}

\section{Numerical simulations}
\label{method}

We run $N$-body simulations of planetary systems
consisting of a host star 
and two planets. 

We use the publicly available 
Bulirsch-Stoer (BS) integration algorithm of MERCURY6.2
with accuracy parameter $\epsilon =10^{-12}$
\citep{chambers99}. We justify the choice of this
algorithm because we are 
mostly interested in the evolution of dynamically active 
systems, where planets experience close encounters, 
and BS handles close encounters better 
than the other integration algorithms in MERCURY6.2.
We simulate the evolution for a maximum time
$t_{\rm max}$ given in units of the initial period 
of the inner planet 
$P_{{\rm in},i}=2\pi\left(Gm_s/a_{{\rm in},i}^3\right)^{-1/2}$,
where $m_s$ is the mass of the central star and
$a_{{\rm in},i}$ is the initial semi-major axis of the inner
planet.
The orbital elements are given in astrocentric coordinates
and the typical conservation of energy and angular momentum 
are better than $\sim10^{-4}$ and $\sim10^{-6}$, respectively.
We ignore the effects from general relativistic precession
and tides in our calculations.

\subsection{Initial conditions and input parameters}

In Table 1, we summarize the input parameters, 
initial conditions, and outcomes of the different 
simulations, which are all described in the following 
subsections.
Our fiducial simulation is {\it 2pl-fiducial}.

The ratios between the mass of the planet  and 
that of the host  for the inner and outer orbits,  
$\mu_{\rm in}$ and $\mu_{\rm out}$, respectively,
are chosen from a uniform distribution in 
$\log$ over the range $[0.1M_J/M_\odot,10M_J/M_\odot]$.
The planets are treated as point masses, not allowing 
for planet-planet collisions. We note that the systems 
that  would have planet collisions are expected to be 
unstable.

In all simulations we start with
a semi-major axis ratio that is uniformly distributed
in $a_{\rm out}/a_{\rm in}\in[3,10]$. 
Thus, we generally exclude from our initial conditions 
the lowest-order mean-motion resonances $p:p+q$, with 
$p=1$ and $q=\{2,3,4\}$ 
($a_{\rm out}/a_{\rm in}=\{1.58,2.08,2.51\}$), 
which can have a strong effect on the dynamics of
the planetary system. Higher-order resonances have a 
weaker effect,
as the strength of the resonant potential is proportional
to $e^q$.

We draw the eccentricities of the inner 
and outer orbits from a uniform distribution in
$[0,0.9]$ and impose an upper limit to the eccentricity of the outer
orbit $e_{\rm out}<1-a_{\rm in}/a_{\rm out}$ to avoid 
a crossing of the initial orbits.
Note that all the orbits very close to this boundary become
unstable and thus do not contribute significant information to 
the derived form of the stability boundary.

For our fiducial simulation {\it 2pl-fiducial}
we initialize the mutual inclinations $i_{\rm m}$ between the inner 
and outer planetary orbits from a Rayleigh distribution
with parameters $\sigma_i=1^\circ$: the corresponding
 mean and median mutual inclinations are 
$1^\circ.25$ and $1^\circ.17$.
In the simulations {\it 2pl-inc-0} and {\it 2pl-inc-20}
we fix $i_{\rm m}=0$ and $i_{\rm m}=20^0$, while in
{\it 2pl-inc-rand} we initialize $i_{\rm m}$ 
from a  uniform distribution in $[0,80^\circ]$.

The arguments of pericenter, the longitudes of 
ascending node,
and the mean anomalies are all drawn from a uniform
distribution in $[0,360^\circ]$.

\subsection{Dynamical outcomes}

We classify the different dynamical outcomes 
into the following categories.
\begin{enumerate}
\item {\it Two planets:} two planets remain in the system
for a time $t_{\rm max}$.
Within this category, we distinguish the systems in which 
the initial semi-major axes of both planets
have changed at a final time $t_{\rm max}$ by less than
$10\%$:
$\left| a_{{\rm in},f}-a_{{\rm in},i}\right|/a_{{\rm in},i}<0.1$
and $\left| a_{{\rm out},f}-a_{{\rm out},i}\right|/a_{{\rm out},i}<0.1$.
These systems have experienced only a small orbital energy 
exchange.
In the complementary category at least one of the planets
has changed its initial semi-major by $10\%$ or more.

\item {\it Ejection:} one planet is ejected from the 
system, which we define to happen
when the planet reaches a distance
from the central star $>100a_{{\rm in},i}$. 
Such planets would almost certainly escape the system
because at this distance the planet is either
in an escape orbit (i.e., eccentricity $\geq1$)
or will most likely soon reach
a escape orbit  by energy
perturbations from the inner planet.

\item {\it Stellar collision:} one planet collides
with the star. 
This is the only scale-dependent outcome because
it depends on our definition of the ratio between
the stellar radius and the initial semi-major axis
$R_\star/a_{{\rm in},i}$.
We use a fiducial conservative value for collisions
of $R_\star/a_{{\rm in},i}=10^{-4}$, equivalent 
to placing the inner planet at $a_{{\rm in},i}=46.5$ AU
for a solar-like star.
We study the effect of larger values of
$R_\star/a_{{\rm in},i}$ in \S\ref{sec:stellar}.
\end{enumerate}

We treat the planets as point masses,
not allowing for collisions between planets. 
However, for two Jupiter-size planets orbiting a Sun-size star 
the ratio $a_{{\rm in},i}/R_J$ in our 
fiducial simulation is $\sim10^5$, which is high
enough that the rate of collisions between planets is expected 
to be very small compared to the rate of ejections
or collisions with the star (e.g., \citealt{PTR14}).

\begin{figure}[h!]
   \centering
  \includegraphics[width=8.4cm]{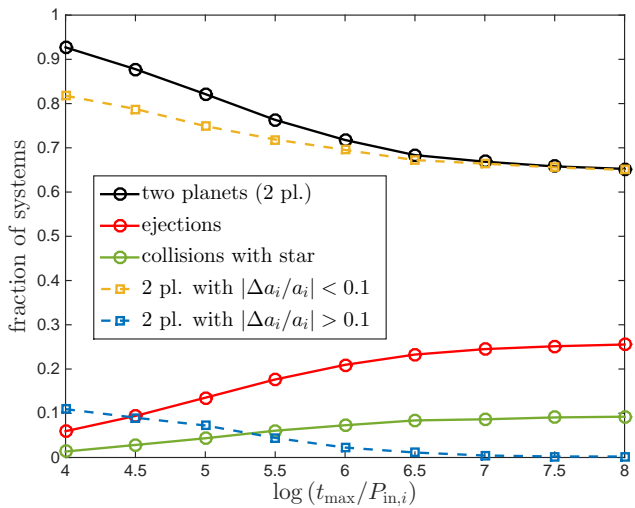}
  \caption{Fraction of systems with different 
  dynamical outcomes as a function of the maximum 
  integration time $t_{\rm max}$ in units of the initial 
  orbital period of the inner planet
  $P_{{\rm in},i}$. 
  The dashed lines indicate the systems
  with two surviving planets for which at least one of the orbits
has either changed its initial semi-major axis by
   $>10\%$ at $t_{\rm max}$ (blue) or not (yellow). 
  }
\label{fig:tmax_outcomes}
\end{figure}   

\subsection{Results}
\label{sec:results_outcomes}

From Table 1, we observe that most systems ($\simeq65\%$) 
in our fiducial simulation {\it 2pl-fiducial} have 
two planets by the end of the simulation.
Within this category over $99\%$
are in secularly stable orbits in the sense that
the planets have experienced only small 
orbital energy variations relative to their initial energies
(the rms $\Delta a/a_i$ of the systems with $|\Delta a|/a_i<0.1$
  is $\simeq0.3\%$, where $a_i$
is the initial semi-major axis).

The second most common outcome ($\simeq26\%$)  is
a system with one planet ejection, followed by a system 
with a stellar collision ($\simeq9\%$).  

The branching ratios into the different dynamical outcomes
depend on various parameters, which we study next.

\subsubsection{Effect of the integration timescale $t_{\rm max}$}

In Figure \ref{fig:tmax_outcomes} and Table 1 we show the evolution 
of the  different outcomes in our fiducial simulation 
as a function of the integration timescale $t_{\rm max}$.
In Table 1 the simulation {\it 2pl-fid-x} corresponds
to {\it 2pl-fiducial}  at $t_{\rm max}=10^xP_{{\rm in,}i}$.

From Figure \ref{fig:tmax_outcomes} and Table 1, we observe that 
the number of systems with two planets 
decreases as a function of time (or $t_{\rm max}$) 
at the expense of increasing the number
of ejections and collisions with the star, as expected.
This decrease is most rapid for $t_{\rm max}<10^6P_{{\rm in,}i}$,
after which time the fraction of systems with two planets
(black line) shows a much slower decrease.
For instance, from Table 1 we see that 
the number of two-planet systems decreases 
by $\simeq12.6\%$ in going from $10^5$ to $10^6 P_{{\rm in,}i}$,
while it does so only by $\simeq2.4\%$  from 
$10^7$ to $10^8 P_{{\rm in,}i}$.

From Figure  \ref{fig:tmax_outcomes} we observe that the fraction
of systems with planets having significant variations in their semi-major
axes ($\left|\Delta a/a_i\right|>0.1$, blue dashed line) decreases rapidly 
from $\simeq11\%$ at $t_{\rm max}=10^4P_{{\rm in,}i}$
to $<0.5\%$ at $t_{\rm max}>10^7P_{{\rm in,}i}$.
Thus, almost all the of systems with two planets that survive for 
more than $10^7P_{{\rm in,}i}$ have experienced small
orbital energy variations relative to their initial values
and might be regarded as secularly stable systems.

In conclusion, our simulations show that there is little 
variation in the branching ratios of the
dynamical outcomes after integrating the systems
for longer than $\sim10^7P_{{\rm in,}i}$. 
After this time the systems with two planets are essentially 
all in secularly stable orbits.

\begin{figure}[h!]
   \centering
  \includegraphics[width=8.4cm]{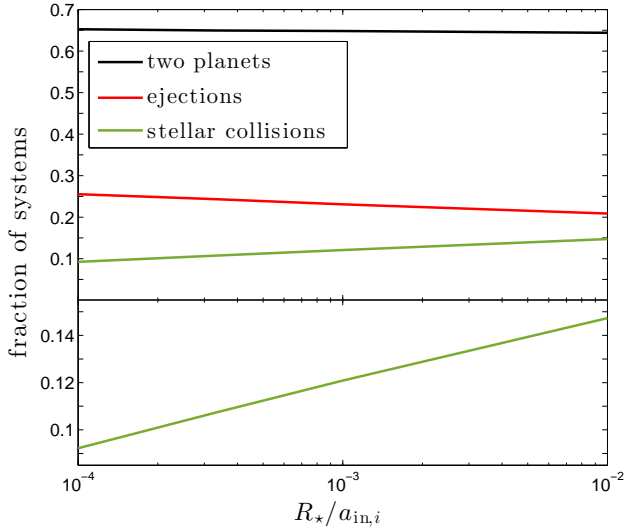}
  \caption{Fraction of systems in {\it 2pl-fiducial}
  with different dynamical outcomes as a function of the 
 the ratio between the stellar radius and the
 initial semi-major axis of the inner planets 
 $R_\star/a_{{\rm in},i}$. The lower panel is a zoom-in 
 of the fraction of systems experiencing stellar
 collisions.}
\label{fig:R_star}
\end{figure} 

\subsubsection{Effect of varying $R_\star/a_{{\rm in},i}$}

In Figure \ref{fig:R_star}, we show the fraction of systems
in {\it 2pl-fiducial} with different outcomes as a function of the ratio between 
the stellar radius and the initial semi-major axis of the inner
planet, $R_\star/a_{{\rm in},i}$.

We observe that the number of collisions with the
star (green lines) increases with $R_\star/a_{{\rm in},i}$,
as expected.
For instance, the fraction of collisions using 
$R_\star/a_{{\rm in},i}=10^{-4}$
($a_{{\rm in},i}\sim50$ AU
for a solar-radius star) is $\simeq9\%$, while
this ratio increases to $\simeq15\%$ for 
$R_\star/a_{{\rm in},i}=10^{-2}$ 
($a_{{\rm in},i}\sim0.5$ AU
for a solar-radius star).

From Figure \ref{fig:R_star}, we observe that the fraction of
systems with two planets (black line) decreases only slightly
($\sim1\%$) by increasing 
$R_\star/a_{{\rm in},i}$ from $10^{-4}$ to $10^{-2}$, while
the fraction of ejections decreases more significantly 
($\sim20\%$) for the same range of $R_\star/a_{{\rm in},i}$.

In summary, varying $R_\star/a_{{\rm in},i}$
mainly affects the ratio between ejection and collisions with the
star, while the fraction of stable and unstable 
(ejections or collisions with the star) remains roughly constant.
We shall use our conservative fiducial value of
$R_\star/a_{{\rm in},i}=10^{-4}$ to determine
the stability boundary in our subsequent analysis.

\begin{figure}[h!]
   \centering
  \includegraphics[width=8.7cm]{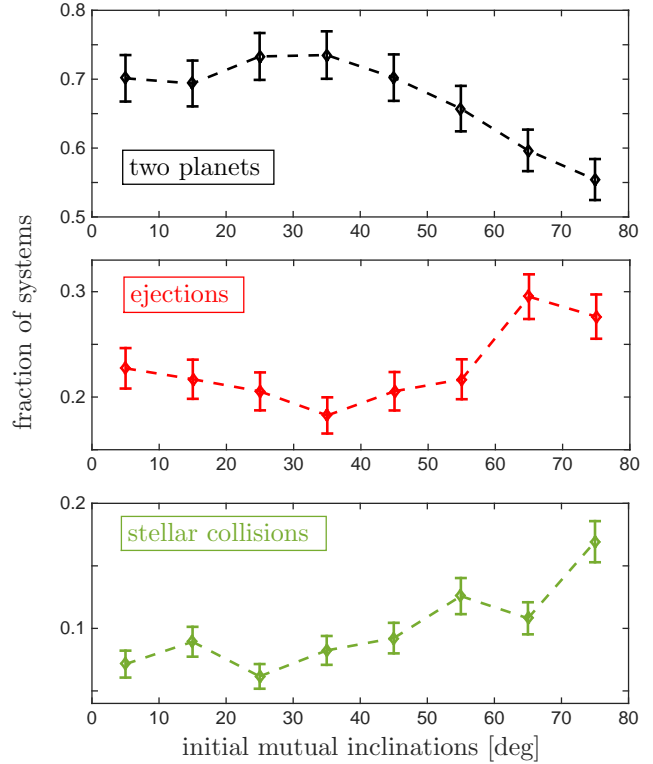}
  \caption{Fraction of systems in {\it 2pl-inc-rand} at
  $t_{\rm max}=10^7 P_{{\rm in},i}$
  as a function of the initial mutual inclination of the two planets.
{\it Upper panel}: systems with two surviving planets.
{\it Middle panel}: systems with one planet ejection.
{\it Lower panel}: systems with one stellar collision.
The error bars indicate the Poisson counting errors for each 
inclination bin.}
\label{fig:hist_inc}
\end{figure} 

\subsection{Effect of the mutual inclination}
\label{sec:itot_effect}

In Figure \ref{fig:hist_inc} we show the fraction of systems
 in {\it 2pl-inc-rand}  with different outcomes for different bins
of the initial mutual inclination $i_{\rm m}$.

The figure shows that the fraction of ejections
decreases from $\simeq0.23$ for $i_{\rm m}<10^\circ$  
to  $\simeq0.16$ for $i_{\rm m}\in[30^\circ, 40^\circ]$.  
This decrease is marginally significant
and might be related to the expected reduction in the 
time at which the planets experience close approaches
when the orbits have higher mutual inclinations.
For the same range of mutual inclination $i_{\rm m}<40^\circ$
the fraction of stellar collisions does not show a clear trend.
However, we observe a statistically significant 
decrease from $\simeq0.09$ at $i_{\rm m}\in[30^\circ, 40^\circ]$ 
to  $\simeq0.06$ for $i_{\rm m}\in[20^\circ, 30^\circ]$.

As we start increasing the mutual inclinations 
from $i_{\rm m}\sim40^\circ$ there is a clear and nearly monotonic
increase in the rate
of both ejections and collisions with the star. 
This behavior might be expected since larger values of
$i_{\rm m}>40^\circ$ can excite Kozai-Lidov eccentricity oscillations
with large amplitudes, which can either decrease the pericenter 
distance to $<R_\star/a_{{\rm in},i}$ producing stellar collisions,
or simply increase the apocenter distance of the inner planet,
promoting close encounters with the outer planet.
As a consequence, the
fraction of systems with two planets decreases significantly 
from $\simeq0.73$ for $i_{\rm m}\in[30^\circ, 40^\circ]$ 
to  $\simeq0.55$ for $i_{\rm m}\in[70^\circ, 80^\circ]$.  

In conclusion, the main effect of increasing the mutual 
inclination from $\sim40^\circ$ is the enhancement
of the rate of ejections and 
collisions with the star.
As we increase the mutual inclinations from $\lesssim10^\circ$ to
$\sim30^\circ-40^\circ$ there is a marginally significant  
decrease in the rate of ejections.

\section{Support Vector Machine (SVM) and
stability boundary}

Our main goal is to find a 
stability boundary that best classifies the different
outcomes and is simple enough (e.g., it has a small 
number of parameters)
to allow for easy interpretation and use.
We shall assess the performance of such a classification 
by its degree of ``completeness,"
defined as the fraction of systems with true 
outcome $X$ that are correctly classified as $X$, or
the ratio between the number of true positives and the
number of true positives plus the number of false
negatives (e.g., \citealt{ivezic}).

We use the Support Vector Machine (SVM) algorithm 
(e.g., \citealt{vapnik96}) to separate
the $i_{\rm 2pl}=1,2,..., N_{\rm 2pl}$ systems with two surviving 
planets from the $i_{\rm ej}=1,2,..,N_{\rm ej}$ systems with planet 
ejections and the $i_{\rm star}=1,2,..,N_{\rm star}$ 
systems with stellar collisions.

We start by defining a set of
parameters $\boldsymbol{\alpha}$ to define a classification
boundary. 
We assume that $\boldsymbol{\alpha}$ is a simple function
of the initial orbital elements $\{a_{\rm out}/a_{\rm in}, e_{\rm in}, 
e_{\rm out}, \Delta\varpi, i_{\rm m}\}$ and the masses
$\{\mu_{\rm in}, \mu_{\rm out}\}$.
For instance, we will define one set of parameters as
$\boldsymbol{\alpha}=[r_{\rm ap}, \mu_{\rm in}^{1/3}]$ with 
$r_{\rm ap}$ defined in Equation (\ref{eq:rap})
and for each system $i=1,2,..,N_{\rm syst}$ we have 
a vector $\boldsymbol{\alpha}_i$.

We separate the classes by a hyperplane
\ba
f(\boldsymbol{\alpha})=\beta_0+\boldsymbol{\beta}\cdot\boldsymbol{\alpha}^t,
\ea
where $\beta_0$ and $\boldsymbol{\beta}$ are constants obtained
using SVM. 
We define the separating function $f(\boldsymbol{\alpha})$
such that $f(\boldsymbol{\alpha})>0$ corresponds to
systems with two planets (that is, stable systems), 
while $f(\boldsymbol{\alpha})<0$ could be either
ejections or collisions with the star. 
We classify only two classes at the time:  
ejections from two surviving
planets in \S\ref{sec:ejections} and stellar collisions
from two surviving planets in \S\ref{sec:stellar}.

For each system 
$i=1,2,..,N_{\rm syst}$ we calculate
$f(\boldsymbol{\alpha}_i)$
and we can formally define the completeness for 
each outcome as:
\ba
f_{\rm 2pl}&=&\frac{ \left\vert f(\boldsymbol{\alpha}_{i_{\rm 2pl}})>0\right\vert}{N_{\rm 2pl}}\\
f_{\rm ej}&=&\frac{ \left\vert f(\boldsymbol{\alpha}_{i_{\rm ej}})<0\right\vert}{N_{\rm ej}}\\
f_{\rm star}&=&\frac{ \left\vert f(\boldsymbol{\alpha}_{i_{\rm star}})<0\right\vert}{N_{\rm star}}
\ea
where $|\cdot|$ is the cardinality of the set of systems and
$f_{\rm 2pl},f_{\rm ej},f_{\rm star}\in[0,1]$. 
Thus, a function that perfectly separates ejections (stellar collisions) 
from two surviving 
planets has $f_{\rm 2pl}=f_{\rm ej}=1$ ($f_{\rm 2pl}=f_{\rm star}=1$), while a 
conservative stability boundary would have $f_{\rm ej}\simeq1$
($f_{\rm star}=1$)
and significantly smaller $f_{\rm 2pl}$.

We train the SVM classifier using
the {\it fitcsvm} package from Matlab 2015a
with standardized variables and a linear Kernel.
For our fiducial simulation, we show the performance of each 
separation (i.e., $f_{\rm 2pl},f_{\rm ej},f_{\rm star}$ in Table 2 and 3)
using the same data as that in the training set. 
We have checked that the completenesses change 
by $\lesssim1\%$ when 
a different set with $\sim$1800 systems
and similar initial conditions
is used to test the performance of the classification.

As discussed in \S\ref{sec:results_outcomes}, the number of
systems with two surviving planets in our
simulations is always larger than the number 
of systems with either ejections or collisions with the star. 
Thus, the SVM algorithm would naturally tend to classify the 
stable systems with a higher completeness than ejections
or collisions with star. 
Since we would like to have a boundary that separates 
each class with similar completeness ($f_{\rm 2pl}\sim f_{\rm ej}$
and $f_{\rm 2pl}\sim f_{\rm star}$), 
we  use a cost matrix in the SVM algorithm such that the
cost of classifying a system into class $X$ if its true class is $Y$
is $N_X/(N_X+N_Y)$. By doing so, we assign a higher penalty
to misclassifying a class with a smaller number 
of systems. 

In practice, this arbitrary procedure works well for defining boundaries 
with similar completenesses and it
mostly changes the offset of $f(\boldsymbol{\alpha})$ 
by a small amount relative to the case with equal misclassification
costs. 
Finally, we note that by artificially promoting a better classification of systems
with either ejections or collisions with the stars (smaller sample) 
at the expense of a poorer classification of stable systems, 
we expect to find a more conservative stability boundary
in the sense that a smaller number of unstable systems are in 
stable regions.

\begin{table*}[ht]
\begin{center}
\caption{Summary of functions $f(\boldsymbol{\alpha})$ found with SVM and from other works
used to separate stable systems from systems with ejections}
\setlength{\tabcolsep}{8pt}
\renewcommand{\arraystretch}{1.7}
\begin{tabular}{c|c|ccc}
\hline
\hline
Simulation&$f(\boldsymbol{\alpha})$ & $f_{\rm 2pl}$ & $f_{\rm ej}$ &$f_{\rm star}$\\
\hline
{\it 2pl-fiducial}&$r_{\rm ap}-1.83$&0.89 & 0.92&0.89\\
{\it 2pl-fiducial}&$r_{\rm ap}+0.2\cos\Delta\varpi - 1.83$&0.89 & 0.92 &0.87\\
{\it 2pl-fiducial}&$r_{\rm ap}-4\mu_{\rm in}^{1/3}-1.40$& 0.91 & 0.95&0.80\\
{\it 2pl-fiducial}&$r_{\rm ap}-0.1\mu_{\rm out}^{1/3}-1.83$&0.89 & 0.92& 0.83\\
{\it 2pl-fiducial}&$r_{\rm ap}-2.7(\mu_{\rm in}+\mu_{\rm out})^{1/3}-1.47$&0.90 & 0.93&0.89\\
{\it 2pl-fiducial}&$r_{\rm ap}-3.8\mu_{\rm in}^{2/7}-1.25$& 0.91 & 0.95&0.79\\
{\it 2pl-fiducial}&$r_{\rm ap}-0.82\mu_{\rm in}^{1/3}(a_{\rm out}/a_{\rm in})-1.27$&0.93 & 0.95&0.69\\
{\it 2pl-fiducial}&$r_{\rm ap}-2.4\mu_{\rm in}^{1/3}(a_{\rm out}/a_{\rm in})^{1/2}-1.15$&0.94 & 0.96&0.75\\
{\it 2pl-fiducial}&$r_{\rm ap}-3.2\mu_{\rm in}^{1/3}(a_{\rm out}/a_{\rm in})^{1/3}-1.15$&0.93 & 0.96&0.74\\
{\it 2pl-fiducial}&$r_{\rm ap}-2.1\mu_{\rm in}^{2/7}(a_{\rm out}/a_{\rm in})^{1/2}-1.03$&0.94 & 0.96&0.75\\
\hline
{\it 2pl-fid-5}&$r_{\rm ap}-2.4\mu_{\rm in}^{1/3}(a_{\rm out}/a_{\rm in})^{1/2}-0.81$&0.99 & 0.96&0.57\\
{\it 2pl-fid-6}&$r_{\rm ap}-2.4\mu_{\rm in}^{1/3}(a_{\rm out}/a_{\rm in})^{1/2}-1.01$&0.97 & 0.96&0.65\\
{\it 2pl-fid-7}&$r_{\rm ap}-2.4\mu_{\rm in}^{1/3}(a_{\rm out}/a_{\rm in})^{1/2}-1.09$&0.95 & 0.96&0.71\\
\hline
{\it 2pl-inc-0}&$r_{\rm ap}-2.4\mu_{\rm in}^{1/3}(a_{\rm out}/a_{\rm in})^{1/2}-1.15$&0.93 & 0.96&0.68\\
{\it 2pl-inc-20}&$r_{\rm ap}-2.4\mu_{\rm in}^{1/3}(a_{\rm out}/a_{\rm in})^{1/2}-1.15$&0.92 & 0.95&0.71\\
{\it 2pl-inc-rand}&$r_{\rm ap}-2.4\mu_{\rm in}^{1/3}(a_{\rm out}/a_{\rm in})^{1/2}-1.15$&0.92 & 0.89&0.57\\
{\it 2pl-inc-rand} ($i_{\rm m}\leq20^\circ$)&$r_{\rm ap}-2.4\mu_{\rm in}^{1/3}(a_{\rm out}/a_{\rm in})^{1/2}-1.15$&0.92 & 0.95&0.71\\
{\it 2pl-inc-rand} ($20^\circ \leq i_{m}\leq40^\circ$)&$r_{\rm ap}-2.4\mu_{\rm in}^{1/3}(a_{\rm out}/a_{\rm in})^{1/2}-1.15$&0.92 & 0.95&0.73\\
{\it 2pl-inc-rand} ($40^\circ \leq i_{\rm m}\leq60^\circ$)&$r_{\rm ap}-2.4\mu_{\rm in}^{1/3}(a_{\rm out}/a_{\rm in})^{1/2}-1.15$&0.92 & 0.90&0.56\\
{\it 2pl-inc-rand} ($60^\circ \leq i_{\rm m}\leq80^\circ$)&$r_{\rm ap}-2.4\mu_{\rm in}^{1/3}(a_{\rm out}/a_{\rm in})^{1/2}-1.15$&0.92 & 0.80&0.41\\
\hline
{\it 2pl-fiducial}&$r_{\rm ap}-Y_{\rm crit}^{\rm EK95}$&0.92 & 0.84& 0.87\\
{\it 2pl-fiducial}&$r_{\rm ap}-Y_{\rm crit}^{\rm MA01}$&0.69 & 0.98 &0.93\\
{\it 2pl-fiducial}&$r_{\rm ap}-Y_{\rm crit}^{\rm Hill}$&0.80 & 0.77 &0.81\\
{\it 2pl-fid-4}&$r_{\rm ap}-Y_{\rm crit}^{\rm Hill}$&0.80 & 0.84 &0.97\\
{\it 2pl-fiducial}&$r_{\rm ap}-Y_{\rm crit}^{\rm GMC13}$&0.63 & 0.99&0.99\\
\hline
\end{tabular}
\end{center}
\label{table:all_sim}
\end{table*}

\subsection{Separation of ejections and two surviving
planets}
\label{sec:ejections}

 We start by separating the systems with two surviving planets
 and from those with ejections because these classes dominate 
 the branching ratios, and we leave the separation of stellar 
 collisions and two planets for \S\ref{sec:stellar}.

\begin{figure*}
   \centering
  \includegraphics[width=18cm]{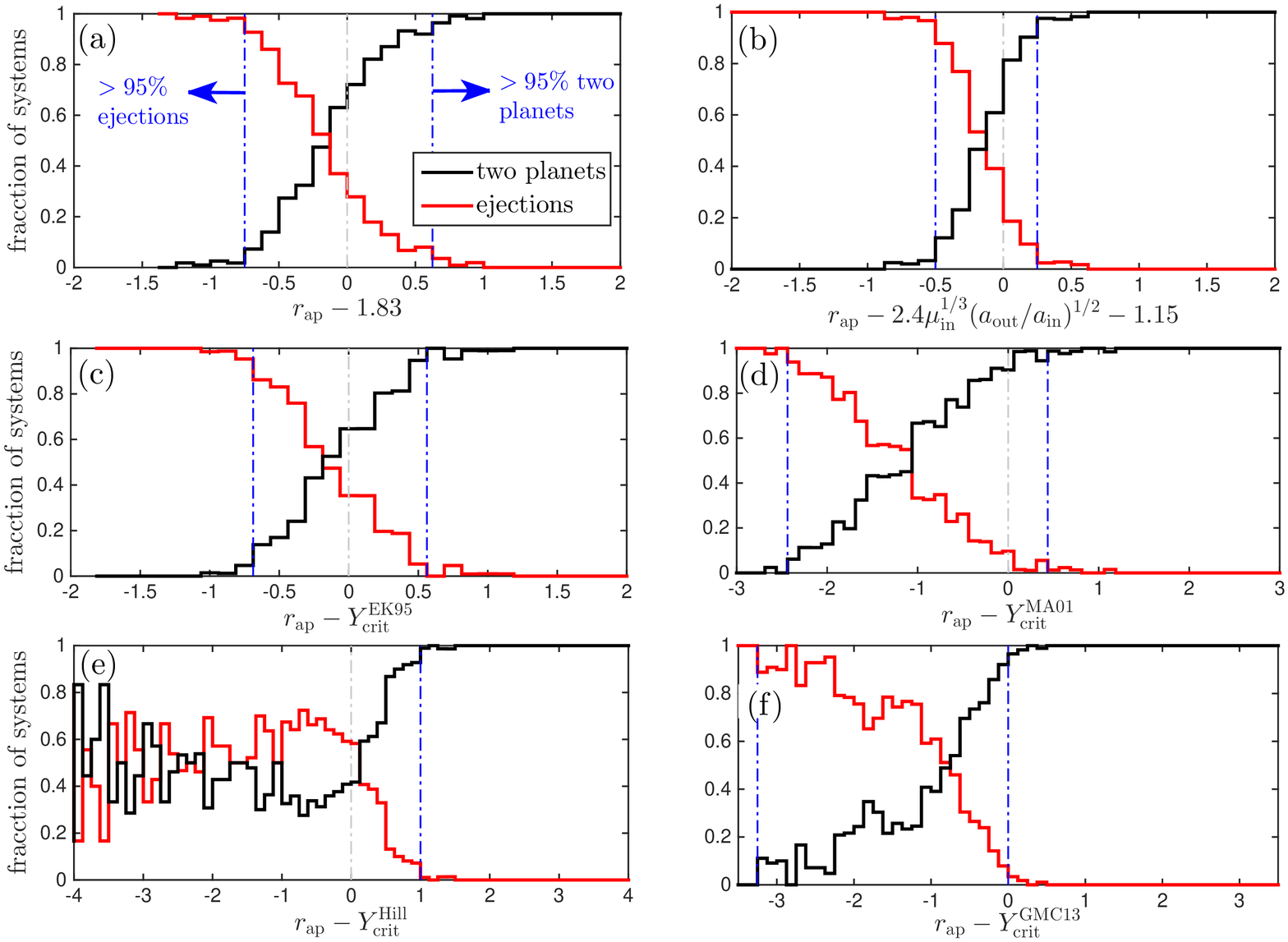}
  \caption{Distribution of the ratio between the number of systems with two 
  surviving planets (solid black line) and ejections (solid red line) and 
  the total number of systems with either two surviving planets or ejections
  (i.e., all systems ignoring collisions with the star) as a function
  of different stability boundaries. The vertical
  dashed-dotted blue lines indicate the regions for which $>95\%$
  of the systems to the left (right) consist of ejections (two planets).
  Panel (a): single-parameter boundary $f=r_{\rm ap}-1.83$ with
  $r_{\rm ap}=a_{\rm out}(1-e_{\rm out})/a_{\rm in}(1+e_{\rm in})$  (see 
  \S\ref{sec:single_alpha}).
Panel (b): two-parameter boundary 
$f=r_{\rm ap}-2.4\mu_{\rm in}^{1/3}(a_{\rm out}/a_{\rm in})^{1/2}-1.15$ (see \S\ref{sec:two_alpha}).
Panel (c): $f=r_{\rm ap}-Y_{\rm crit}^{\rm EK95}$ 
from \citet{EK95} in Equation (\ref{eq:EK95}).
Panel (d): $f=r_{\rm ap}-Y_{\rm crit}^{\rm MA01}$ 
from \citet{MA01}
in Equation (\ref{eq:MA01}).
Panel (e): $f=r_{\rm ap}-Y_{\rm crit}^{\rm Hill}$ from 
\citet{gladman93} in Equation (\ref{eq:Y_hill}).
Panel (f): $f=r_{\rm ap}-Y_{\rm crit}^{\rm GMC13}$ from 
\citet{GMC13} in Equation (\ref{eq:GMC13}).
Note that the horizontal axes of panels (d), (e), and (f)
are different from those in panels (a) through (c).\\
  }
\label{fig:hist_all}
\end{figure*}

 In Table 2 we show the completeness 
 for different separating functions $f(\boldsymbol{\alpha})$ 
 in different simulations (see Table 1).
 We also include a set of previously proposed stability 
 boundaries from Equations (\ref{eq:EK95}), (\ref{eq:MA01}),
(\ref{eq:Y_hill}), and (\ref{eq:GMC13})
 in \S\ref{sec:previous_stab}.
Similarly, in Figure \ref{fig:hist_all} we show the distribution
of the ratio between the number of systems with two surviving
  planets (solid black line) and ejections (red black line) and 
  the total number of systems with either two planets or ejections
  for the stability boundaries above.
  The best criteria are those with values of $f$ closest to unity.

\subsubsection{A single parameter stability boundary:
$f(\alpha)=\beta_0+\beta_1\alpha$}
\label{sec:single_alpha}

We start by constructing a stability boundary that 
only depends  on one parameter, using our fiducial
simulation {\it 2pl-fiducial}. 
We choose the parameter
to depend on only the initial orbital elements and ignore
the masses because without the orbital elements the masses
cannot predict the fate of a planetary system.

From Table 1, we observe that by setting $\alpha=r_{\rm ap}$ we
obtain the function $f(\alpha)=r_{\rm ap}-1.83$ in our fiducial simulation.
For this boundary we find  completenesses
of $f_{\rm 2pl}\simeq0.86$ and $f_{\rm ej}\simeq0.87$.
Recall that by setting $\Delta\varpi=180^\circ$, 
the parameter $r_{\rm ap}$ becomes a measure of the minimum 
distance  $d_{\rm min}$
between two non-crossing coplanar orbits and 
$d_{\rm min}=a_{\rm in}(1+e_{\rm in})(r_{\rm ap}-1)$.
Thus, the boundary can be rewritten as 
$d_{\rm min}=0.83\cdot a_{\rm in}(1+e_{\rm in})$;
in words, the boundary classifies a system as stable if initially its 
orbits have a minimum distance that
is at least  $\simeq83\%$ of the apocenter distance of the inner 
planet.

One could calculate the initial 
minimum distance of the two ellipses for arbitrary values
$\Delta\varpi$ and construct a stability boundary using a
more precise measure of the closest approaches between
the planets. However, this approach has a few shortcomings:
\begin{enumerate}
\item the relative orientation of the orbits seems to have little
effect of the performance of the stability boundary. In Table 2
we show that adding the extra parameter $\cos(\Delta\varpi)$
does not increase the values of $f_{\rm 2pl}$ and $f_{\rm ej}$.
\item The resulting expression is too complicated to be of any 
practical use.
\end{enumerate}

Based on the arguments above, we will ignore the dependence 
on the initial relative apsidal angles $\Delta\varpi$ in our
subsequent analysis.  
We are aware that in a case-by-case basis, the relative orientation 
can certainly make a difference for the stability boundary
(see the discussion in \S\ref{sec:miss} and \citealt{GMC13}).

In summary, we argue that the best single-parameter  
stability boundary is $r_{\rm ap}=1.83$ because of its simple functional
form and the relatively high values of completenesses it 
achieves,  $f_{\rm 2pl}\simeq0.86$ and $f_{\rm ej}\simeq0.87$.

 \subsubsection{A two-parameter stability boundary:
$f(\boldsymbol{\alpha})=\beta_0+\beta_1\alpha_1+\beta_2\alpha_2$}
\label{sec:two_alpha}
 
Based on our findings above that the best single parameter to describe the
stability boundary is $r_{\rm ap}$, we fix $\alpha_1\equiv r_{\rm ap}$
and  vary the functional form of $\alpha_2$ to search 
for a two-parameter stability boundary that best separates stable systems from 
those with ejections in {\it 2pl-fiducial}. 
 
We start by including the dependence on the planet-to-star mass ratios
$\mu_{\rm in}$ and $\mu_{\rm out}$ in $f(\boldsymbol{\alpha})$.
Motivated by the dependence of Hill's stability criterion on
the planet-to-star mass ratios, we test the performance of the 
following parameters:
$\alpha_2=\mu_{\rm in}^{1/3}$, $\mu_{\rm out}^{1/3}$, and
$(\mu_{\rm in}+\mu_{\rm out})^{1/3}$.
From Table 2, we observe that the parameter that performs the best among
these choices is $\alpha_2=\mu_{\rm in}^{1/3}$ because it reaches
the highest completenesses, $f_{\rm 2pl}\simeq0.89$ and 
$f_{\rm ej}\simeq0.92$ compared to $f_{\rm 2pl}\simeq0.86$ and 
$f_{\rm ej}\simeq0.87$ for the one-parameter boundary.
Also, we notice that incorporating the parameter 
$\mu_{\rm out}^{1/3}$ provides almost no improvement in the completeness
relative to the single-parameter boundary $f(\alpha)=r_{\rm ap}-1.83$ 
and SVM assigns a small multiplicative coefficient 
of $\beta_2\simeq-0.1$.
Finally, a stability boundary using the 
parameter $(\mu_{\rm in}+\mu_{\rm out})^{1/3}$ marginally improves 
the performance of the boundary relative to the single-parameter
expression, but it performs worse than simply using $\mu_{\rm in}^{1/3}$.

We have tried different power laws of the form $\mu_{\rm in}^{\theta}$
and found for $\theta=\{1/2,1/3,2/7,1/4\}$ the highest completenesses
are reached with either $1/3\simeq0.333$ or $2/7\simeq0.286$ (see Table 2). 
The performance is significantly 
(marginally) worse using $\theta=1/2$ ($\theta=1/4$).
Based on these results our method
does not distinguish between the performance of a 
boundary with $\theta=1/3$, which is expected from
Hill stability \citep{gladman93} and a boundary 
with $\theta=2/7$, expected from resonance 
overlap \citep{wisdom80}. 
Note that \citet{MW12} extended the work by 
\citep{wisdom80} to planets with non-zero eccentricities
and found a different power law with $\theta=1/5$, which 
is not favored by our experiments relative to either $\theta=1/3$ or
$\theta=2/7$.

We experimented with various simple functional forms 
$g(a_{\rm in},a_{\rm out},e_{\rm in},e_{\rm out})$
in $\alpha_2=\mu_{\rm in}^{1/3}g$ and found that 
by setting $g=(a_{\rm in}/a_{\rm out})^{\nu}$ with $\nu>0$ tends to 
increase the completeness relative to $g=1$.
In Table 1, we show the results of the stability boundaries
for $\nu=\{1,1/2,1/3\}$ and found the best results
with $\nu=1/2$.
With this index the stability boundary is
\ba
f=r_{\rm ap}-2.4\mu_{\rm in}^{1/3}(a_{\rm out}/a_{\rm in})^{1/2}-1.15,
\label{eq:ap_1}
\ea
while for a $\alpha_2=\mu_{\rm in}^{2/7}g$
we find: 
\ba
f=r_{\rm ap}-2.1\mu_{\rm in}^{2/7}(a_{\rm out}/a_{\rm in})^{1/2}-1.03.
\label{eq:ap_2}
\ea

The boundaries in Equations (\ref{eq:ap_1}) and (\ref{eq:ap_2})
yield the highest  completenesses in our heuristic search,
$f_{\rm 2pl}=0.94$ and $f_{\rm ej}=0.95$. 
We decided to stop here because the
completenesses reached are close to unity. 
Also, note that we have experimented by adding an extra
parameter $\alpha_3$ with various functional forms
and found only a marginal increase in the
completenesses ($\sim1\%$), which are 
at the expense of a more complicated expression 
of $f$.

In summary, we have found that the mass of the outer
planet carries no information regarding the stability 
against planetary ejections 
in our simulations and we only need to know the 
mass of the inner planet. 
We have found two stability boundaries with
different power-laws for the inner planet-to-star
$\mu_{\rm in}$ (Eqs. [\ref{eq:ap_1}] and [\ref{eq:ap_2}]), 
which perform the best based both on the 
completeness they reach 
when separating ejections and stable systems
and on their simplicity.

\begin{figure}[h!]
   \centering
  \includegraphics[width=8.4cm]{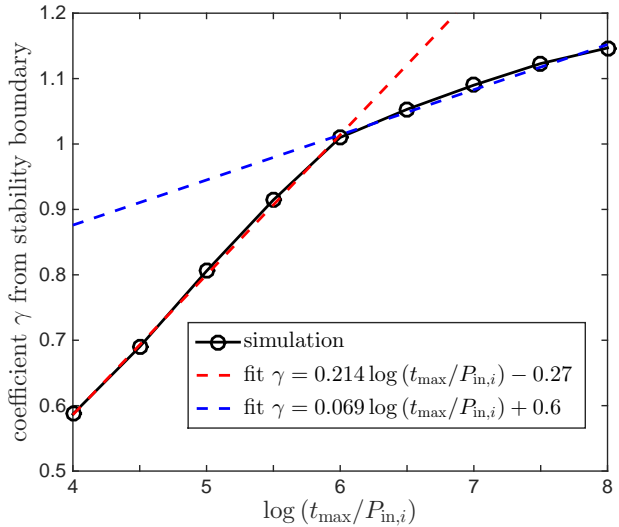}
  \caption{Coefficient $\gamma$ from the stability boundary
   $r_{\rm ap}=2.4\mu_{\rm in}^{1/3}(a_{\rm out}/a_{\rm in})^{1/2}+\gamma$
  as a function of the maximum integration 
  time $t_{\rm max}$ in units of the initial 
  orbital period of the inner planet
  $P_{{\rm in},i}$. The red and blue dashed lines indicate
  a linear fit for $\log\left(t_{\rm max}/P_{{\rm in},i}\right)\in[4,6]$ and
  $\log\left(t_{\rm max}/P_{{\rm in},i}\right)\in[6,8]$, respectively}.
\label{fig:gamma}
\end{figure}  

\subsubsection{Stability boundary  and the 
maximum integration timescale.}

We have found that a simple function that separates stable
from unstable systems in {\it 2pl-fiducial} is given by 
Equation (\ref{eq:ap_1}).
Here we study the effect of the maximum integration time on this
stability boundary.
To do so we write the stability boundary as
$r_{\rm ap}=2.4\mu_{\rm in}^{1/3}(a_{\rm out}/a_{\rm in})^{1/2}+\gamma$
and see how $\gamma$ varies with $t_{\rm max}$.

In Figure \ref{fig:gamma}, we show our results 
for $\gamma$  as a function of  $t_{\rm max}$.
We show similar results in Table 2,
labeled as {\it 2pl-fid-x} (e.g., $\gamma=1.01$
for $t_{\rm max}=10^6P_{{\rm in},i}$).
From Table 2, we observe that the functional form above 
can well separate the stable systems from 
ejections  ($f_{\rm 2pl},f_{\rm ej}\gtrsim0.95$) just by changing 
$\gamma$. 

From Figure \ref{fig:gamma}, we observe that the required value of
$\gamma$ increases  monotonically 
from $\simeq0.6$ for $\log\left(t_{\rm max}/P_{{\rm in},i}\right)=4$ to
$1.15$ for $\log\left(t_{\rm max}/P_{{\rm in},i}\right)=8$.
This increase is expected because
systems with larger $r_{\rm ap}$ (fixing the masses
and semi-major axes) should become unstable later.
We also observe that the coefficient $\gamma$ increases
 more than $\sim3$ times more rapidly with time from 
$\log\left(t_{\rm max}/P_{{\rm in},i}\right)=4$ to 
$\log\left(t_{\rm max}/P_{{\rm in},i}\right)=6$
than for $\log\left(t_{\rm max}/P_{{\rm in},i}\right)\geq6$
(see linear fits).

From Figure \ref{fig:gamma}, we fit the evolution of $\gamma$ for
$\log\left(t_{\rm max}/P_{{\rm in},i}\right)\geq6$ and find
that the stability boundary is given by
\ba
r_{\rm ap}=2.4\mu_{\rm in}^{1/3}\left(\frac{a_{\rm out}}{a_{\rm in}}\right)^{1/2}
+0.069\log\left(\frac{t_{\rm max}}{P_{{\rm in},i}}\right) + 0.6.\nonumber\\
\label{eq:stab_time}
\ea

This stability boundary is valid for 
$t_{\rm max}/P_{{\rm in},i}=10^6-10^8$ 
and given the slow variation of $\gamma$ with
time, it suggests that only a small fraction of stable systems 
in {\it 2pl-fiducial}  can become unstable in longer
timescales.

\subsubsection{Misclassified systems}
\label{sec:miss}

Some of the stable systems ($\simeq6\%$) are 
classified as ejections because they initially satisfy
$r_{\rm ap}<2.4\mu_{\rm out}^{1/3}(a_{\rm out}/a_{\rm in})^{1/2}+1.15$.
These systems tend to start with relatively aligned orbits: 
$\simeq50\%$ ($\simeq82\%$) start with $\cos\Delta\varpi>0.8$
($\cos\Delta\varpi>0$).
We checked that in some extreme cases, the system starts 
with $r_{\rm ap}<0$ and 
avoids orbit crossing by starting with $\cos\Delta\varpi\sim1$ 
and engaging in a secular resonance. 

Similarly, some of the systems with ejections ($\simeq4\%$) are 
classified as stable because they initially satisfy 
$r_{\rm ap}>2.4\mu_{\rm out}^{1/3}(a_{\rm out}/a_{\rm in})^{1/2}+1.15$.
These systems tend to start with relatively misaligned orbits: 
$\simeq40\%$ ($\simeq76\%$) start with $\cos\Delta\varpi>0.8$
($\cos\Delta\varpi>0$).

By adding the extra parameter $\cos\Delta\varpi$ to
our stability boundary we find 
$r_{\rm ap}=2.4\mu_{\rm out}^{1/3}(a_{\rm out}/a_{\rm in})^{1/2}+
0.2\cos\Delta\varpi +1.1$ and 
the completenesses increase only marginally 
from $f_{\rm 2pl}\simeq0.94$ and $f_{\rm ej}\simeq0.96$
to $f_{\rm 2pl}\simeq0.94$ and $f_{\rm ej}\simeq0.97$.

In conclusion, some of the misclassification might be
explained by the initial relative orientation of the 
ellipses since the orbits that start with more aligned (misaligned)
pericenters tend to be more stable (unstable). 
These results are consistent with the 
claims by \citet{GMC13}. 
However, the overall effect of $\Delta\varpi$  only 
marginally improves the performance from our simpler stability
boundary.

\begin{table}[ht]
\begin{center}
\caption{Summary of functions $f(\boldsymbol{\alpha})$ found with SVM used to separate 
stable systems from systems with stellar collisions}
\setlength{\tabcolsep}{3pt}
\renewcommand{\arraystretch}{1.7}
\begin{tabular}{c|c|ccc}
\hline
\hline
Simulation&$f(\boldsymbol{\alpha})$ & $f_{\rm 2pl}$ & $f_{\rm ej}$ &$f_{\rm star}$\\
\hline
{\it 2pl-fiducial}&$r_{\rm ap}-1.83$& 0.89 & 0.91&0.89\\
{\it 2pl-fiducial}&$r_{\rm ap}+0.47\mu_{\rm in}^{1/3}-1.93$& 0.89 & 0.91&0.89\\
{\it 2pl-fiducial}&$r_{\rm ap}-3.4\mu_{\rm out}^{1/3}-1.45$& 0.91 & 0.87&0.91\\
{\it 2pl-fiducial}&$r_{\rm ap}-2.7(\mu_{\rm in}+\mu_{\rm out})^{1/3}-1.58$&0.90 & 0.93&0.89\\
{\it 2pl-fiducial}&$r_{\rm ap}-0.9\mu_{\rm out}^{1/3}(a_{\rm out}/a_{\rm in})-1.18$& 0.93 & 0.83&0.92\\
{\it 2pl-fiducial}&$r_{\rm ap}-2.4\mu_{\rm out}^{1/3}(a_{\rm out}/a_{\rm in})^{1/2}-1.15$& 0.92 & 0.83&0.92\\
\hline
\end{tabular}
\end{center}
\label{table:fid_coll}
\end{table}

\begin{figure*}
   \centering
  \includegraphics[width=18cm]{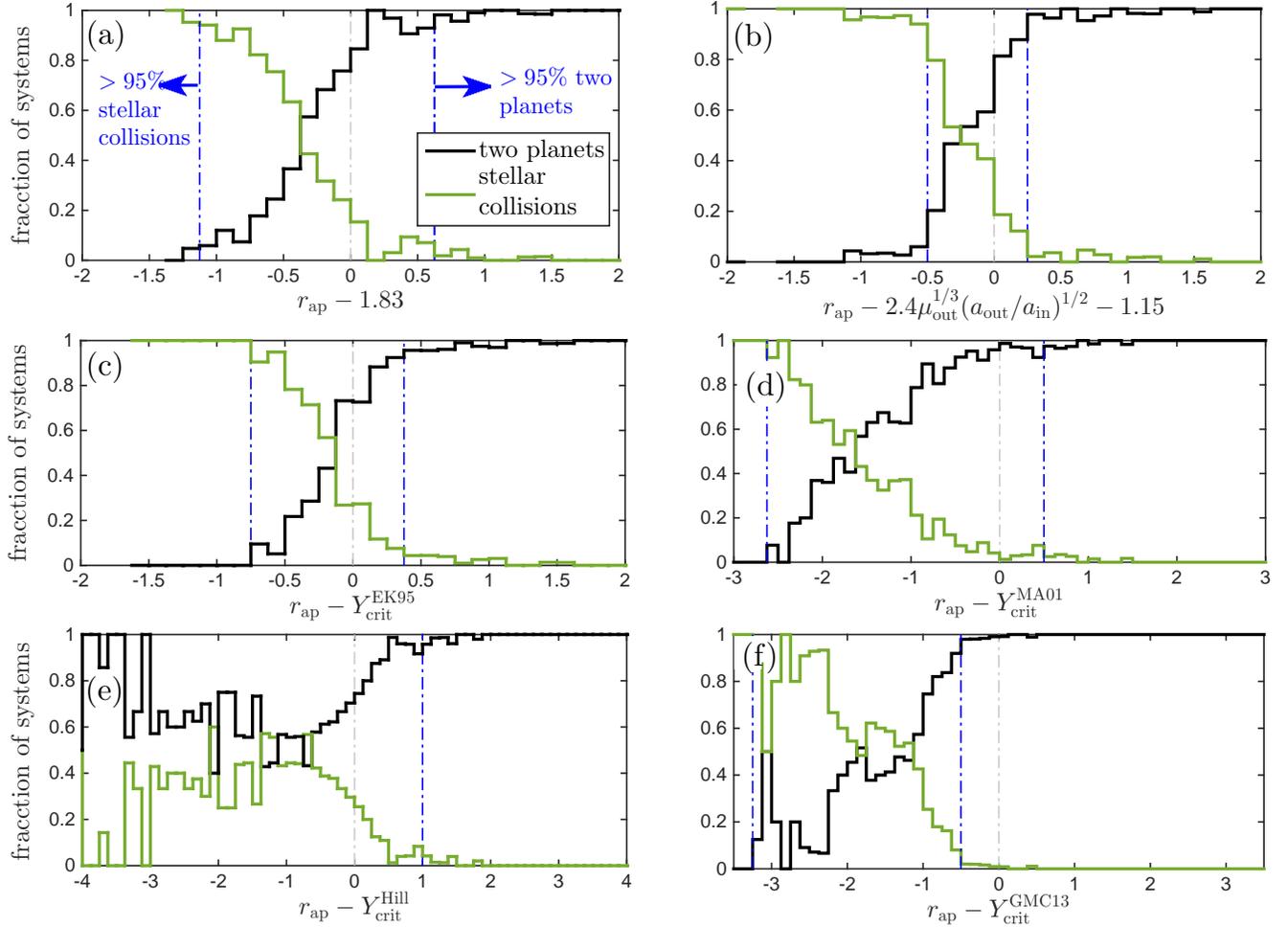}
  \caption{Distribution of the ratio between the number systems with two 
  surviving planets (solid black line) and stellar collisions (solid green line) and 
  the total number of systems with either two surviving planets or stellar collisions (i.e., all systems ignoring ejections) as a function
  of different stability boundaries. The vertical
  dashed-dotted blue lines indicate the regions for which $>95\%$
  of the systems to the left (right) consist of stellar collisions (two planets).
  Panel (a): single-parameter boundary $f=r_{\rm ap}-1.83$ with
  $r_{\rm ap}=a_{\rm out}(1-e_{\rm out})/a_{\rm in}(1+e_{\rm in})$  (see 
  \S\ref{sec:stellar}).
Panel (b): two-parameter boundary $f=r_{\rm ap}-2.4\mu_{\rm out}^{1/3}(a_{\rm out}/a_{\rm in})^{1/2}-1.15$ (see \S\ref{sec:stellar}).
Panel (c): $f=r_{\rm ap}-Y_{\rm crit}^{\rm EK95}$ 
from \citet{EK95} in Equation (\ref{eq:EK95}).
Panel (d): $f=r_{\rm ap}-Y_{\rm crit}^{\rm MA01}$ 
from \citet{MA01}
in Equation (\ref{eq:MA01}).
Panel (e): $f=r_{\rm ap}-Y_{\rm crit}^{\rm Hill}$ from 
\citet{gladman93} in Equation (\ref{eq:Y_hill}).
Panel (f): $f=r_{\rm ap}-Y_{\rm crit}^{\rm GMC13}$ from 
\citet{GMC13} in Equation (\ref{eq:GMC13}).
Note that the horizontal axes of panels (d), (e), and (f)
are different from those in panels (a) through (c).\\
  }
\label{fig:hist_all_coll}
\end{figure*}

\subsection{Separation of stellar collisions and two surviving
planets}
\label{sec:stellar}

Following the same procedure as in \S\S\ref{sec:single_alpha} 
and \ref{sec:two_alpha}, we search for a stability boundary 
that separates systems that experience stellar collisions from
systems with two surviving planets in our fiducial simulation
{\it 2pl-fiducial}. 

In Table 3, we show our results for some separating
functions found using SVM and their corresponding  
completenesses.
Similarly, in Figure \ref{fig:hist_all_coll} we show the distribution 
of the ratio between the number of systems with two planets 
(solid black line) and collisions (solid red line) and the total number
 of systems with either two planets or collisions for different
 stability boundaries, including those in
 Equations (\ref{eq:EK95})-(\ref{eq:Y_hill}), and (\ref{eq:GMC13})
from \S\ref{sec:previous_stab}.

From Table 3, we observe that the single-parameter 
boundary using $r_{\rm ap}$ is given by 
$f(\alpha)=r_{\rm ap}-1.83$, which is identical to that
found in  \S\ref{sec:single_alpha} for separating
ejections from systems with two surviving planets.
The completenesses using this function are
 $f_{\rm 2pl}=0.89$ and  $f_{\rm star}=0.89$.

As in \S\ref{sec:two_alpha}, we
include the dependence on the planet-to-star mass ratios
$\mu_{\rm in}$ and $\mu_{\rm out}$ in $f(\boldsymbol{\alpha})$, and
test the performance of the following parameters:
$\alpha_2=\mu_{\rm in}^{1/3}$, $\mu_{\rm out}^{1/3}$, and
$(\mu_{\rm in}+\mu_{\rm out})^{1/3}$.
From Table 3, we observe that the boundary with
$\alpha_2=\mu_{\rm in}^{1/3}$ does not improve the performance 
relative to the single-parameter boundary (the completenesses
are the same). 

By setting $\alpha_2=(\mu_{\rm in}+\mu_{\rm out})^{1/3}$ 
we observe that there is a slight improvement 
since  $f_{\rm 2pl}$ increases from 0.89 in the single-parameter
boundary to 0.9 and the resulting separating boundary is very similar
to the one found for separating ejections from stable systems
(see Table 2).
Finally, by setting $\alpha_2=\mu_{\rm out}^{1/3}$ we find that the
performance improves more significantly and the completenesses
are  $f_{\rm 2pl}=f_{\rm star}=0.91$.
We tried other functional forms for the mass ratios and observed
no improvements relative to $\alpha_2=\mu_{\rm out}^{1/3}$.

Similar to our procedure in \S\ref{sec:two_alpha}, we add dependence
on the semi-major axis  ratio $a_{\rm out}/a_{\rm in}$.
We find that the best separation is reached by setting 
$\alpha_2=\mu_{\rm out}^{1/3}(a_{\rm out}/a_{\rm in})$ 
with completenesses of $f_{\rm 2pl}=0.93$
and $f_{\rm star}=0.92$ (see Table 3).
A slightly worse separation ($f_{\rm 2pl}=f_{\rm star}=0.92$) is
reached with $\alpha_2=\mu_{\rm out}^{1/3}(a_{\rm out}/a_{\rm in})^{1/2}$,
where the function reads
\ba
f=r_{\rm ap}-2.4\mu_{\rm out}^{1/3}(a_{\rm out}/a_{\rm in})^{1/2}-1.15.
\label{eq:ap_coll}
\ea
This function is identical to that in Equation 
(\ref{eq:ap_1}), found
to separate ejections from
stable systems if we change $\mu_{\rm out}$ for $\mu_{\rm in}$.
This surprising  result suggests that the long-term stability of the
system against either ejections or collisions might only
depend on $\max(\mu_{\rm in},\mu_{\rm out})$.
Then, if an unstable system has $\mu_{\rm in}>\mu_{\rm out}$, the
most likely outcome is an ejection, while a collision with the host star 
is slightly more likely if  $\mu_{\rm in}<\mu_{\rm out}$.
Motivated by these findings we favor the separating function in 
Equation (\ref{eq:ap_coll}) over
$r_{\rm ap}-0.9\mu_{\rm out}^{1/3}(a_{\rm out}/a_{\rm in})-1.18$.
Finally, we have experimented with different exponents $\nu$
in $\alpha=\mu_{\rm out}^{1/3}(a_{\rm out}/a_{\rm in})^{\nu}$
and found no improvement relative to $\nu=1/2$.

In summary, the mass of the outer (and not the inner) planet carries 
most of the information about the systems that collide with the star.
We found that a good stability boundary for separating collisions
from stable systems is given by Equation (\ref{eq:ap_coll}), which 
is identical to the one found
for separating ejections from stable systems when changing
$\mu_{\rm out}$ for $\mu_{\rm in}$.

\begin{figure}
   \centering
  \includegraphics[width=8.5cm]{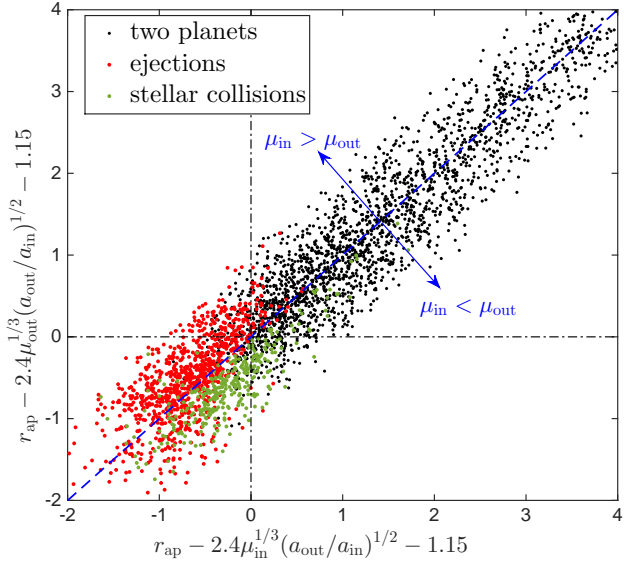}
  \caption{Boundary found for separating stellar collisions from 
  stable systems from Equation (\ref{eq:ap_coll})
  as a function of the boundary found for separating ejections
  from stable systems Equation (\ref{eq:ap_1}) for the different outcomes
  in {\it 2pl-fiducial}.
  The diagonal dot-dashed blue line indicate boundary in which 
  both boundaries cross (i.e., $\mu_{\rm in}=\mu_{\rm out}$).
}
\label{fig:coll_ej}
\end{figure}  

\subsection{A criterion for stability against either
ejections or stellar collisions}

In \S\ref{sec:ejections} and \S\ref{sec:stellar} we
have found criteria for separating systems with
ejections from systems with two surviving
planets and systems with
stellar collisions from systems with two surviving
planets, respectively.
In Figure \ref{fig:coll_ej} we show these two criteria
by plotting the stability boundary
against ejections in Equation (\ref{eq:ap_1}) versus
the stability boundary against collisions with the star
in Equation (\ref{eq:ap_coll}) for the different outcomes
in {\it 2pl-fiducial}.
We note that these criteria differ only on the whether 
the mass of the inner and the outer is used.

We observe that most systems in regions
in the positive quadrant (Equations
[\ref{eq:ap_1}] and [\ref{eq:ap_coll}] are both positive)
are stable (very few red and green dots).
This result suggest that we can combine
Equations (\ref{eq:ap_1}) and (\ref{eq:ap_coll})
to define a stability condition against 
either ejections or collisions with the star as:
\ba
r_{\rm ap}>2.4\left[\max\{\mu_{\rm in},\mu_{\rm out}\}\right]^{1/3}
(a_{\rm out}/a_{\rm in})^{1/2}+1.15.
\label{eq:ap_max}
\ea
This criterion yields completenesses of 
$f_{\rm 2pl}\simeq0.9$ and $f_{\rm ej+star}\simeq0.95$.
Since $f_{\rm 2pl}<f_{\rm ej+star}$ this criterion is somewhat 
conservative in the sense that a smaller number of unstable 
systems are in stable regions relative to the number of stable systems
in unstable regions.

Similarly, we observe from Figure \ref{fig:coll_ej} that
most systems in the negative quadrant are unstable
and by using the minimum instead the maximum in
Equation (\ref{eq:ap_max}) we get 
$f_{\rm 2pl}\simeq0.81$ and $f_{\rm ej+star}\simeq0.97$.
Moreover, within the unstable systems we observe that
almost all of the collisions with the star ($\simeq95\%$) are
in regions where $\mu_{\rm in}<\mu_{\rm out}$, while
most ($\simeq72\%$) of the systems with ejections 
have $\mu_{\rm in}>\mu_{\rm out}$.
Since we have an overall higher rate of ejections than
stellar collisions, we find that both rates are comparable 
in regions where $\mu_{\rm in}<\mu_{\rm out}$:
$\simeq45\%$ and $\simeq55\%$ of the unstable
systems undergo ejections and collisions
with the star, respectively.

In summary, we combine our previous results in
Equation (\ref{eq:ap_max}) to propose a stability 
boundary against either ejections or collisions
with the star.
Systems that are unstable and have
$\mu_{\rm in}>\mu_{\rm out}$ will most likely
undergo a planet ejection, while the
systems that have $\mu_{\rm in}<\mu_{\rm out}$
will have a similar rate of ejections and
stellar collisions, with the latter being slightly higher.

\begin{figure}
   \centering
  \includegraphics[width=8.5cm]{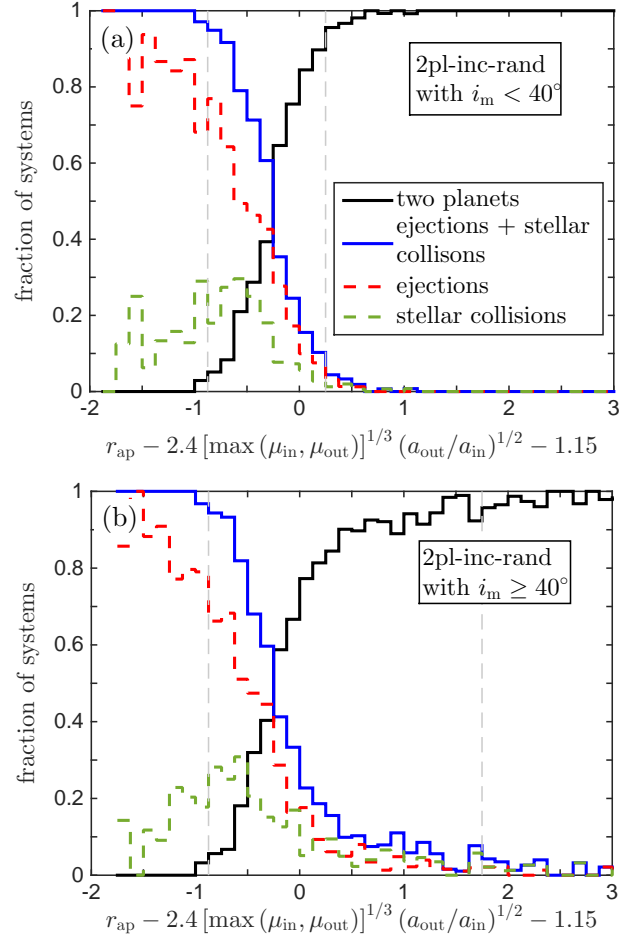}
  \caption{Fraction of systems with two planets (solid
  black line), either ejections or collisions with the star
  (solid blue line), ejections (dashed red line),
  and collisions with the star (dashed green line)
  in the simulation  {\it 2pl-inc-rand} as function of
  the stability boundary in Equation (\ref{eq:ap_max}).
  {\it Panel a:} systems with mutual inclinations
  $i_{\rm m}<40^\circ$.
    {\it Panel b:} systems with mutual inclinations
  $i_{\rm m}<40^\circ$.
  The vertical dashed gray lines indicate the regions
   for which $>95\%$
  of the systems to the left (right) consist of either
  ejections or collisions with the star (two planets).
}
\label{fig:inc_e}
\end{figure}  

\subsection{Effect of mutual inclinations on
the stability boundary}

From Table 2 we observe that the stability boundary
against ejections in Equation (\ref{eq:ap_1}) found using
{\it 2pl-fiducial} performs relatively well in {\it 2pl-inc-rand}
($f_{\rm 2pl}\simeq0.92$ and $f_{\rm 2pl}\simeq0.89$), which
has a random distribution of the mutual inclination 
in $i_{\rm m}\rm [0,80^\circ]$.

By taking different bins of $i_{\rm m}$ in {\it 2pl-inc-rand} we
find that the performance of the boundary in Equation (\ref{eq:ap_1})
is the same ($f_{\rm 2pl}\simeq0.92$ and $f_{\rm 2pl}\simeq0.95$)
for the systems starting with
$i_{\rm m}<20^\circ$ and $i_{\rm m}\in[20^\circ,40^\circ]$.
Similarly, the performance of this boundary in the coplanar
case ($i_{\rm m}=0$) {\it 2pl-inc-0} and in the simulation 
{\it 2pl-inc-20} with $i_{\rm m}=20^\circ$ is almost the same.
Thus, our stability boundary against ejections performs well
for mutual inclinations $i_{\rm m}\lesssim40^\circ$.

As we increase the initial mutual inclinations in {\it 2pl-inc-rand}
we find that $f_{\rm ej}$ drops from $\simeq0.95$ for 
$i_{\rm m}<40^\circ$ to 0.9 and 0.8 for  
$i_{\rm m}\in[40^\circ,60^\circ]$ and $i_{\rm m}\in[60^\circ,80^\circ]$,
respectively. On the contrary, $f_{\rm 2pl}$ remains equal to 
$\simeq0.92$ for all bins in mutual inclinations.
As discussed in \S\ref{sec:itot_effect}, this behavior might be expected 
since larger values of $i_{\rm m}>40^\circ$ can excite 
Kozai-Lidov
eccentricity oscillations, which can promote close
encounters with 
the outer planet and produce ejections in regions that would be
long-term stable for $i_{\rm m}\lesssim40^\circ$.

In Figure \ref{fig:inc_e}, we show the fraction of systems with 
different outcomes 
in our simulation {\it 2pl-inc-rand} for $i_{\rm m}<40^\circ$
in panel (a) and $i_{\rm m}\geq40^\circ$ in panel (b)
 as a function of the stability criterion against either
ejections or stellar collisions in Equation (\ref{eq:ap_max}).
From panel (a) we observe that there in only a small
fraction of systems with either ejections or collisions 
with the star for  
$r_{\rm ap}>2.4\left[\max\{\mu_{\rm in},\mu_{\rm out}\}\right]^{1/3}
(a_{\rm out}/a_{\rm in})^{1/2}+1.15$ and 
$f_{\rm ej}\simeq f_{\rm star}\simeq f_{\rm star+ej}\simeq0.96$
for $i_{\rm m}<40^\circ$.
From panel (b) we observe that this fraction of unstable
systems in stable regions increases for $i_{\rm m}<40^\circ$ 
and the completenesses decrease significantly:
 $f_{\rm ej}\simeq0.87$,  $f_{\rm star}\simeq0.67$,
 and $f_{\rm star+ej}\simeq0.8$.
Since $f_{\rm star}$ is significantly lower than $f_{\rm ej}$ 
we conclude that the performance of our stability criterion
worsens mostly at the expense of having collisions with the star
in regions classified as stable.
This effect is observed in Figure \ref{fig:inc_e} as an increase
in the tail with positive value of Equation (\ref{eq:ap_max}) 
of the distribution of collisions with the star (green dashed line)
in panel (b) relative to panel (a).

In summary, our stability criterion in Equation (\ref{eq:ap_max}) 
performs well for mutual inclinations $i_{\rm m}\lesssim40^\circ$. 
For higher mutual inclinations the Kozai-Lidov mechanism
produces a significant fraction 
of unstable systems in regions classified as stable and
our criterion becomes a poor predictor for long-term stability.
Our results seem consistent with a previous study by 
\citet{geor13}, which shows that the mutual inclination has very
little effect on the stability boundary for $i_{\rm m}\in[0,40^\circ]$.

\section{Discussion}
\label{sec:discussion}

The main results of this paper are a set of new
stability boundaries for separating systems that become
unstable against ejections and collisions with 
the star from systems that retain their
two planets with small orbital energy variations
(Equations [\ref{eq:ap_1}] and [\ref{eq:ap_coll}]).
In particular, we propose 
that hierarchical two-planet systems are long-term
stable if they satisfy the condition in Equation 
(\ref{eq:ap_max}).

Additionally, we find that our stability boundary:
\begin{enumerate}

\item  performs significantly better
than other previously proposed criteria 
(see completenesses in Table 2
and 3, and Figures \ref{fig:hist_all} and \ref{fig:hist_all_coll});

\item performs well for all mutual inclinations $i_{\rm m}\lesssim40^\circ$;

\item and changes slowly with the maximum
integration timescale as $\propto 0.07\log(t_{\rm max}/P_{\rm in})$ 
for $t_{\rm max}/P_{\rm in}=10^6-10^8$, while it does so $\sim3$
times more rapidly for $t_{\rm max}/P_{\rm in}=10^4-10^6$
(see Figure \ref{fig:gamma} and Equation [\ref{eq:stab_time}]);

\end{enumerate}

The fate of the unstable systems depends mostly  
on the planetary masses. Most systems with 
$\mu_{\rm in}>\mu_{\rm out}$ lead to ejections, while 
for $\mu_{\rm in}<\mu_{\rm out}$ there is a slightly higher 
number of collisions with the star than ejections.

In what follows we discuss some of the consequences
of our findings in the context of other works
and the observations.

\subsection{Performance of other stability criteria}
\label{sec:performance}

In Table 2 we show the completenesses
of the different stability criteria discussed in 
\S\ref{sec:previous_stab}
that were reached in our fiducial simulation {\it 2pl-fiducial}.
Similarly, panels (b) to (f)  in Figures \ref{fig:hist_all} and \ref{fig:hist_all_coll}
show the fraction of outcomes for each stability boundary.
In these figures
the performance is best when the outcomes have
the sharpest transition from 0 to 1
(a perfect separation leads to two step-functions).

First, our simulations show that the Hill stability criterion (Equation
[\ref{eq:Y_hill}] for coplanar systems) is a poor indicator of the dynamical stability of 
two-planet systems because it achieves relatively low
completenesses
($f_{\rm 2pl} \sim f_{\rm ej}\sim f_{\rm star}\sim0.8$). 
For comparison, even the single-parameter boundary 
$r_{\rm ap}=1.83$ performs significantly better
($f_{\rm 2pl} \sim f_{\rm ej}\sim f_{\rm star}\sim0.9$).
Moreover, an important fraction of the systems that are 
classified as Hill unstable 
are actually long-term stable (see the solid black lines 
in panel (e) of Figures \ref{fig:hist_all} and \ref{fig:hist_all_coll}
with $r_{\rm ap}<Y_{\rm crit}^{\rm Hill}$). 
\footnote{Consistent with the definition of Hill stability, we have 
checked that in all the unstable systems with
$r_{\rm ap}>Y_{\rm crit}^{\rm Hill}$
it is the outer (inner) planet the one that is ejected (collides with 
the star).
Otherwise the planets would have had orbit crossing
events.}

Second, we observe that both the criteria by \citet{MA01}
and \citet{GMC13} are rather conservative because they 
have $f_{\rm ej},f_{\rm star}>0.93$ and $f_{\rm 2pl}<0.7$.
Thus, the systems satisfying these criteria are expected
to be long-term stable, but those systems that do not satisfy this
condition are not necessarily expected to be unstable.

Finally, we observe that the empirical stability boundary 
by \citet{EK95} performs the best among the previously 
proposed criteria. From Table 1, we observe that 
$f_{\rm 2pl}\simeq0.92$, $f_{\rm ej}\simeq0.84$,
and $f_{\rm star}\simeq0.87$, which are comparable
to those obtained from our
 one-parameter criterion $r_{\rm ap}=1.83$,
but significantly lower than our two-parameter
boundaries in Equations (\ref{eq:ap_1})
and (\ref{eq:ap_coll})

In summary, the stability boundary by 
\citet{EK95} performs the best among the previously 
proposed criteria, while those by
\citet{MA01} and \citet{GMC13} are too conservative.
The Hill stability criterion has poor performance 
and  provides very little useful information
regarding the fate of the Hill unstable systems.

\subsection{Relation to other works with more
than two planets}

Our results are strictly valid only for two-planet systems.
However, some of our main findings can still provide
useful information regarding the 
 long-term stability in systems with
more than two planets.

First, we have found that the stability boundary depends
on the eccentricities only through 
$r_{\rm ap}=a_{\rm out}(1-e_{\rm out})/a_{\rm in}(1+e_{\rm in})$,
which means that the relevant quantity to describe the 
stability is the distance between 
the pericenter of the outer planet and the apocenter of 
the inner planet.
Moreover, we show that the relative orientation 
of the ellipses plays only a minor role in separating a
stable from unstable systems (see \S\ref{sec:miss}).
Consistent with our results, the recent experiments by 
\citet{PW15} show a similar dependence on the
stability boundary in systems with seven planets.
In their study the relevant quantity is the 
distance between the pericenter of the outer planet and the 
apocenter of the inner planet 
for all the adjacent planets.

Second, we find that the stability boundary 
changes $\sim 3$ times more slowly with the maximum integration
time for $t_{\rm max}/P_{{\rm in},i}<10^6$ than in the 
range $t_{\rm max}/P_{{\rm in},i}=10^6-10^8$
(see Figure \ref{fig:gamma}).
We observe a similar behavior in the numerical experiments
by \citet{SL09}, with more than two planets (see Figure 1 therein)
and by \citet{CFMS2008} with three planets (see Figure 29 therein).
There, the slope of the separation between adjacent planets
required for stability as a function of $t_{\rm max}/P_{{\rm in},i}$ 
decreases significantly after
$\sim10^6$ orbits of the innermost planet in the system.

Previous studies generally 
parameterized the spacing required for
stability in units of the mutual Hill radii $R_{\rm H}$
as $K\equiv (a_{\rm out}-a_{\rm in})/R_{\rm H}=a+b \log(t_{\rm max})$
 (e.g., \citealt{chambers96,SL09}), which makes it
hard to make a direct quantitative comparison with
our results in  Equation (\ref{eq:stab_time}). 
 However, we can approximate our stability boundary in 
 Equation (\ref{eq:stab_time}) for the limit of small eccentricities 
 (or $a_{\rm out}-a_{\rm in}\ll a_{\rm out}$) and equal-mass planets 
 ($\mu_{\rm in}=\mu_{\rm out}$) to write
  $K\propto \tilde{b} [3/(2\mu_{\rm in})]^{1/3} \log(t_{\rm max})$, where
$\tilde{b} $ is the coefficient we have
  obtained from our simulations and the Support Vector
  Machine algorithm.
  From our fits in Figure \ref{fig:gamma}, we find
 $\tilde{b} =0.021$ for $t_{\rm max}/P_{{\rm in},i}=10^4-10^6$
and $\tilde{b} =0.069$ for $t_{\rm max}/P_{{\rm in},i}=10^6-10^8$. 
Since our simulations have an average mass ratios
$\bar\mu_{\rm in}=\bar\mu_{\rm out}=10^{-3}$, we derive
$K\propto2.4 \log(t_{\rm max})$ for $t_{\rm max}/P_{{\rm in},i}=10^4-10^6$ and 
$K\propto0.79 \log(t_{\rm max})$ for $t_{\rm max}/P_{{\rm in},i}=10^6-10^8$.

We notice that the slope of $0.79$ that we found for 
$t_{\rm max}/P_{{\rm in},i}=10^6-10^8$ falls in the 
range of $b\sim0.7-1.3$ that was found
in previous studies by \citet{SL09}, \citet{funk10}, and \citet{PW15}.
Our results also show that it is not possible to fit the 
boundary with a single linear fit in  $\log(t_{\rm max})$
since the slope $b$ decreases as a function of time.
This expectation is consistent with 
previous studies that predict lower values of $b$
as the integration time increases $t_{\rm max}$: 
\citet{funk10} predict $b\sim1.3$ for $t_{\rm max}/P_{{\rm in},i}=10^4-10^7$,
\citet{SL09} predict $b\sim1$ for $t_{\rm max}/P_{{\rm in},i}\lesssim10^8$,
and
\citet{PW15} predict $b\simeq0.7$ for $t_{\rm max}/P_{{\rm in},i}=10^7-10^9$
 
In summary, our results show
that the stability boundary depends on the
eccentricities only through the distance between the
orbits and  logarithmically on the maximum integration time, 
with a shallower slope for longer times.
These results  are qualitatively 
consistent with other stability studies with
more than two planets.

\begin{figure}
   \centering
  \includegraphics[width=8.5cm]{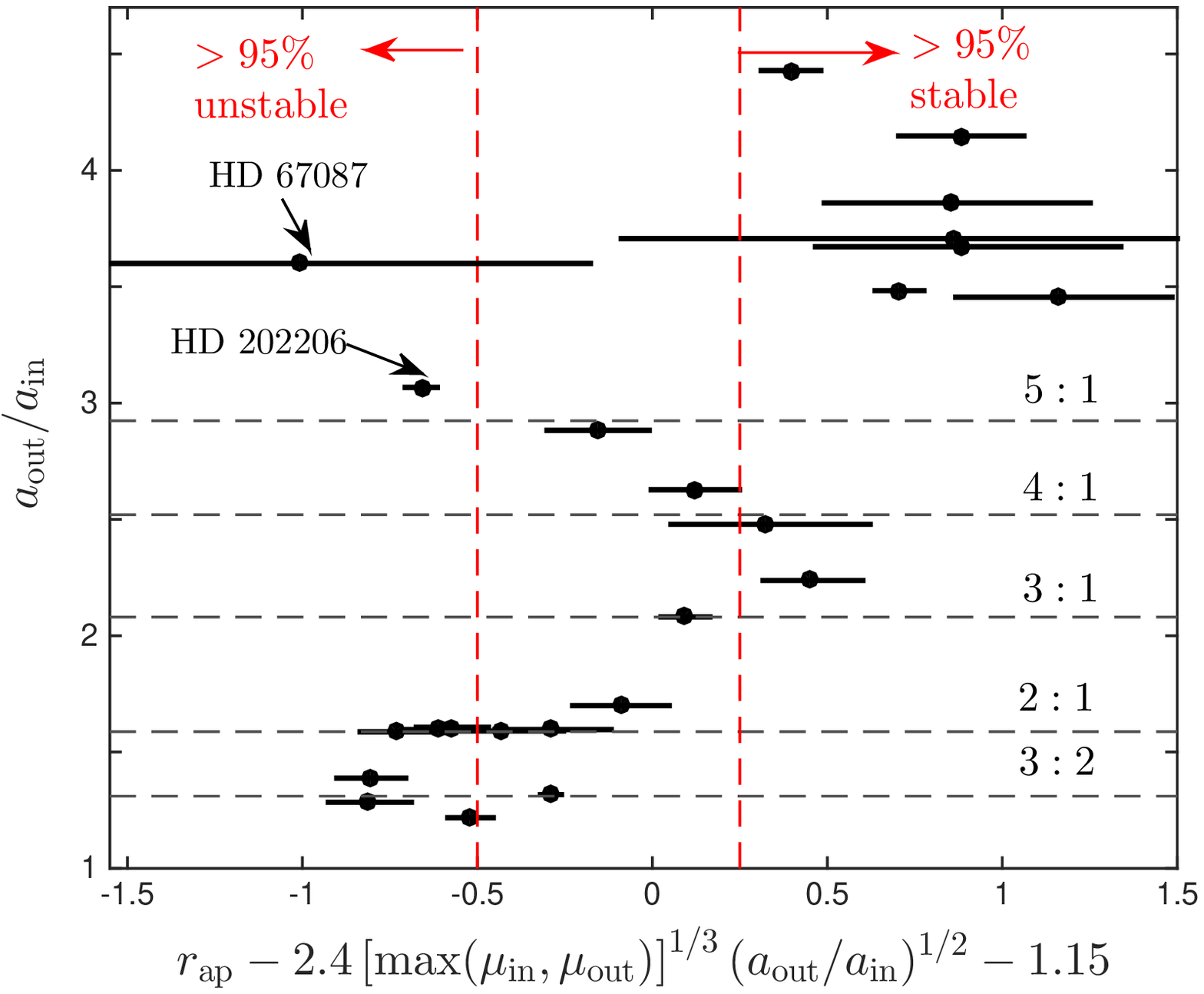}
  \caption{Stability boundary in Equation (\ref{eq:ap_max})
  as a function of the semi-major axis ratio
  $a_{\rm out}/a_{\rm in}$ for a sample of two-planet
  systems discovered by radial velocity surveys
  with  $a_{\rm out}/a_{\rm in}<5$ 
  (from \citealt{wright11} and HD 67087 from
  \citealt{Harakawa15}). 
  The error bars only consider the errors in the eccentricities
  and we use the minimum planet masses to calculate
  $\mu_{\rm in}$ and   $\mu_{\rm out}$. 
  The vertical red dashed lines indicate the regions 
  for which $>95\%$ of the systems to the left (right) 
  are expected to be unstable (stable) according 
  to the stability criterion. The horizontal dashed lines
  indicate the position of the strongest mean-motion
  resonances.\\}
\label{fig:exo}
\end{figure}  

\subsection{Application to observed planetary systems}

  The results from our fiducial simulation {\it 2pl-fiducial} show that
    (see panel (b)  in Figures \ref{fig:hist_all} and
  \ref{fig:hist_all_coll}):
\begin{enumerate}
\item with probability $>0.95$ a system is unstable
if  $r_{\rm ap}<2.4\left[\max(\mu_{\rm in},\mu_{\rm out})\right]^{1/3}+0.6$,
with ejections occurring for $\mu_{\rm in}\geq\mu_{\rm out}$
and either ejection or collisions with the star 
for $\mu_{\rm in}<\mu_{\rm out}$;
\item and with probability $>0.95$ a system is stable against either
ejections or collisions with the star
if $r_{\rm ap}>2.4\left[\max(\mu_{\rm in},\mu_{\rm out})\right]^{1/3}(a_{\rm out}/a_{\rm in})^{1/2}+1.4$.
\end{enumerate}

In Figure \ref{fig:exo} we show the stability boundary 
in Equation \ref{eq:ap_max}
for a sample of two-planet systems discovered by
radial velocity surveys.
We also display the regions 
  for which $>95\%$ of the systems to the left (right) 
  are expected to be unstable (stable) according 
  to the criterion above.

 We observe that some systems are expected to be 
 unstable according to our results. In particular,
 there are 3  and 4 systems around the 
 $2:1$ and $3:2$ that are consistent with being left to the 
dashed  vertical, respectively. 
These results might seem to contradict the validity
of our stability constraints. However, our results apply
to more widely-spaced systems with $a_{\rm out}/a_{\rm in}>3$, 
where we avoid the effect from these first-order 
mean-motion resonances  that can promote the
long-term stability of the system.
  
A more curious case is the two-planet system HD 202206
because it has $a_{\rm out}/a_{\rm in}=3.1$
and our results should apply to this range of  $a_{\rm out}/a_{\rm in}$.
As discussed by \citet{correia05} and 
\citet{couetdic10} such a system 
is indeed unstable 
for the best three-body fit of the RV measurements. 
However, there are stable coplanar solutions provided that the 
system is in a $5:1$ mean-motion resonance.

Recently, \citet{Harakawa15} discovered the
the planetary system HD 67087, 
which contains two planets with minimum masses of 
$\mu_{\rm in}\sim0.002$ and $\mu_{\rm out}\sim0.004$
and  orbital elements $a_{\rm out}/a_{\rm in}\simeq3.6^{+0.24}_{-0.24}$,
$e_{\rm in}=0.17^{+0.07}_{-0.07}$, and 
$e_{\rm out}=0.76^{+0.17}_{-0.24}$.
This system is particularly interesting because 
our stability criterion indicates
that the systems should be unstable  unless 
the value of $e_{\rm out}$ is in the lower end 
of its error measurement (see the error bar in Figure 9). 
This result suggests that the dynamical stability of this system
should be further investigated, including the possibility 
of non-coplanar configurations of the orbits that can lead to 
more stable solutions.

In conclusion, all of the observed systems 
(with the exception of HD 67087) that are likely to be 
unstable according to our criterion seem to be protected 
by a mean-motion resonance. The effect of mean-motion
resonances does not play a significant
role in our calculations because we have excluded 
the lowest-order mean-motion resonances 
$p:p+q$ with $p>0$ and $q=\{1,2,3\}$ 
($a_{\rm out}/a_{\rm in}=\{1.58,2.08,2.51\}$) from
our calculations.

Finally, our results 
can be used to put constraints
on the orbital elements of potential planets in systems with 
RV trends or poorly constrained RV measurements.
In what follows, we give one worked example where we apply 
our stability boundary.

\subsubsection{A worked example: constraints on the 
eccentricity of KOI-1299c.}

KOI-1299 is a giant star harboring at least two giant planets
(e.g., \citealt{ciceri15,ortiz15,quinn15}).
The planetary system is in a hierarchical configuration with
$a_{\rm out}/a_{\rm in}\simeq 4$ and the inner 
planet (KOI-1299b) is in an eccentric orbit $e_{\rm in}\simeq0.5$.
The planet-to-mass ratios are
$\mu_{\rm in}\simeq0.004$ and $\mu_{\rm out}\geq0.0018$. 

Using these parameters and assuming that the planets 
have relatively low mutual inclinations ($i_{\rm m}\lesssim40^\circ$)
and $\mu_{\rm in}>\mu_{\rm out}$, 
our stability constraint above 
implies that the system is unstable against ejections with 
probability $>95\%$ 
if the outer planet has an eccentricity of $e_{\rm out}\gtrsim0.5$. 
Therefore, we conclude that with high probability 
that the eccentricity of KOI-1299c is $e_c\lesssim0.5$.
This upper limit is useful in this example because
the RV measurements by \citet{quinn15} yield 
$e_c=0.64_{-0.13}^{+0.14}$ and 
the error bar can be shrunk by using our stability 
constraint.
Consistently, the authors have indeed studied the
stability of this system and concluded that the system 
can be stable in a coplanar configuration for 
$\sim6\times10^6$ orbits of the inner planet only if $e_c\lesssim0.55$,
which then allowed
them to fit the orbital parameters to much higher accuracy,
finding $e_c=0.498_{-0.059}^{+0.029}$. 

It might be surprising that the upper limit found by \citet{quinn15} is 
less constraining than the one we found with our stability boundary.
However, these authors study the stability of the system surveying 
a much more constrained region of parameter space, confining
the orbits to be almost apsidally aligned, which allows
for stable orbits with higher values of $e_c$ compared to random
orientation of the orbits as we have assumed in our
simulations (see \S\ref{sec:miss}).

We repeated the analysis above by using all the others
stability boundaries shown in Figure \ref{fig:hist_all} to 
determine an upper limit to $e_c$. 
The single-parameter boundary requires $e_c\lesssim0.59$,
while the boundary from \citet{EK95} in Equation (\ref{eq:EK95})
$e_c\lesssim0.57$.
All other stability boundaries that we have tested here 
(panels (d), (e), and (f)  in Figure \ref{fig:hist_all}) do not 
provide a useful constraint as they only demand
$e_c\leq1$ for ejections not occur with probability $>0.95$.

In summary, our stability constraint places a strong constraint
on the eccentricity of KOI-1299c, which is consistent with
the stability analysis of \citet{quinn15} for this system.
All other previously proposed stability boundaries, except that of 
\citet{EK95}, do not place a useful constraint to the eccentricity of
KOI-1299c.

\subsection{Stellar evolution and white dwarf
pollution}

From Table 1, we note that the ratio between the number of
stellar collisions and ejections is $\simeq0.36$, meaning
that  $\simeq27\%$ of the unstable systems reach
distances $<R_\odot$ if the inner planet starts $a_{{\rm in},i}=46.5$ AU.
This fraction increases up to $\simeq41\%$ by placing the planet at
$a_{{\rm in},i}=0.465$ AU (see Figure \ref{fig:R_star}) .
Since a Jupiter-like planet orbiting a $0.6M_\odot$ white dwarf
is expected to be disrupted in a highly eccentric orbit 
if it reaches a distance 
$\lesssim3R_\odot$ \citep{GRL11}, we expect that unstable
hierarchical two-planet systems can often lead to tidal disruptions.

Note that most ($\simeq95\%$) stellar collisions start
with $\mu_{\rm in}<\mu_{\rm out}$ (see Figure \ref{fig:coll_ej}).
Also, we observe that  the ratio between the number stellar
collisions and the number of ejections  in our fiducial simulation 
is $\lesssim0.1$ for $\mu_{\rm out}/\mu_{\rm in}\lesssim1.5$
and it reaches values of $\sim1-3$ for 
$\mu_{\rm out}/\mu_{\rm in}\sim2-6$.
These results are qualitatively consistent with the increase in the
ratio between the number of planets undergoing a 
close approach with the star  and the number of ejections
from $\simeq0.03$ for equal-mass planets
to $\simeq0.12-0.16$ for planetary-mass ratios of $\simeq2.3-3$
(randomly assigning the more massive planet as the inner one)
observed by \citet{FR08}.
However, we observe that the overall rate of collisions with the star
relative to ejections can be several times higher
in our simulations for two initially eccentric planets
relative to the simulations by  \citet{FR08} for two
planets in initially circular orbits.

Equation (\ref{eq:stab_time}) shows that 
as the planetary system ages the
our stability boundary becomes more stringent, 
allowing orbits with relatively larger separations (larger 
$r_{\rm ap}$) to become unstable.
However, the dependence is only logarithmic 
and the boundary moves only by $\sim7\%$ percent 
per order magnitude difference 
in the evolution time.
Thus, by extrapolating this result to timescales 
$>10^8P_{\rm in}$ one would expect only a small effect
in the stability of planetary systems.

A more pronounced effect from the aging of the planetary 
system is likely to come from mass loss of the host star
(e.g., \citealt{DS02}). Typical white dwarfs have masses
that are a few times lower than their main-sequence
progenitors and therefore the mass ratios 
$\mu_{\rm in}$ and $\mu_{\rm out}$ are expected
to increase by the same factor, while keeping 
$a_{\rm out}/a_{\rm in}$ fixed (see, 
\citealt{veras13,mustill14,VG15}). 
This effect is expected to destabilize the  systems close
to our stability boundary in Equation (\ref{eq:ap_max}).

In summary, unstable two-planet systems in an initially
 hierarchical configuration can lead to a significant 
 number of collisions with the star relative to the number of
 ejections, which might contribute to
 the pollution of white dwarfs as a result of stellar 
 mass loss. The number of collisions with the star (or tidal
 disruptions) can 
 be higher than the number of ejections for
 $\mu_{\rm out}/\mu_{\rm in}\sim2-6$.

\section{Conclusions}

We run a large number of long-term numerical
integrations to study
the fates of two-planet systems in
hierarchical configurations with arbitrary eccentricities
and mutual inclinations. 

Using the Support Vector Machine algorithm to separate 
different fates of our simulated systems, we find 
that initially nearly coplanar systems remain long-term stable 
for $a_{\rm out}(1-e_{\rm out})/[a_{\rm in}(1+e_{\rm in})]>2.4
\left[\max(\mu_{\rm in},\mu_{\rm out})\right]^{1/3}(a_{\rm out}/a_{\rm in})^{1/2}+1.15$.
Systems that do not satisfy this condition by a margin
of $\gtrsim0.5$ are expected to be unstable,
mostly leading to planet ejections if $\mu_{\rm in}>\mu_{\rm out}$, 
while slightly favoring collisions with the star for $\mu_{\rm in}<\mu_{\rm out}$.

We show that our proposed stability boundary 
performs significantly better than previously proposed stability 
criteria (\citealt{EK95}, \citealt{MA01}, and Hill stability) 
for mutual  inclinations $\lesssim 40^\circ$.

\acknowledgements 
 I acknowledge support from the 
CONICYT Bicentennial  Becas Chile fellowship. 
I am indebted to my PhD advisor Scott Tremaine, who has 
critically read and commented on various versions 
of this paper.
I thank Jose Garmilla for helping me with technical issues 
regarding the Support
Vector Machine algorithm and Dimitri Veras for
helping me with technical aspects of the Mercury integrator.
I am grateful to Dimitri Veras, Renu Malhotra, 
and Katherine Deck for enlightening
discussions and comments.
All simulations were carried out using computers 
supported by the Princeton Institute of Computational 
Science and Engineering.
This research has made use of the Exoplanet Orbit Database
and the Exoplanet Data Explorer at exoplanets.org.


\end{document}